\documentclass{article}

\RequirePackage[OT1]{fontenc}
\RequirePackage{amsthm,amsmath}
\RequirePackage[round, authoryear]{natbib}
\RequirePackage[colorlinks,linkcolor=blue,citecolor=blue,urlcolor=blue]{hyperref}
\usepackage{times}
\usepackage{bm}
\usepackage{amsmath}
\usepackage{amsfonts}
\usepackage{amsthm} 
\usepackage{amssymb}
\usepackage{amstext}
\usepackage{bbm}
\usepackage{mathtools}
\usepackage{graphicx}
\usepackage{adjustbox}
\graphicspath{ {./plots/} }
\usepackage[mathscr]{euscript}
\usepackage{tabularx}
\usepackage{mathrsfs}
\usepackage{enumerate} 
\usepackage{cases} 
\usepackage{float} 
\usepackage{caption}
\usepackage{subcaption}
\usepackage{comment}
\usepackage{wrapfig}
\usepackage[english]{babel}
\usepackage[utf8]{inputenc}
\usepackage[colorinlistoftodos]{todonotes}
\usepackage[T1]{fontenc}
\usepackage{amsbsy} 
\usepackage{physics} 
\usepackage{upgreek} 
\usepackage{diagbox} 
\usepackage{rotating}
\usepackage{authblk}

\theoremstyle{plain}

\newtheorem{theorem}{Theorem}[section]
\newtheorem{lemma}[theorem]{Lemma}

\theoremstyle{definition}

\theoremstyle{remark}

\def\Mcal{{\mathcal{M}}}
\def\Vcal{{\mathcal{V}}}
\def\Xcal{{\mathcal{X}}}

\def\Pr{{\text {pr}}}

\def \bsm {\boldsymbol}
 
\def \balpha {\boldsymbol{\alpha}}
\def \bpsi {\boldsymbol{\psi}}
\def \bPsi {\boldsymbol{\Psi}}

\def \bLambda {\boldsymbol{\Lambda}}

\def \bbeta {\boldsymbol{\beta}}

\def \bq {\boldsymbol{\mathrm{q}}}

\def \bp {\boldsymbol{p}} 
\def \bM {\boldsymbol{M}}
\def \bP {\boldsymbol{P}}
\def \bQ {\boldsymbol{Q}}

\def \bI {\boldsymbol{I}}
\def \drm {\mbox{{\tiny{\sc drm}}}}

\def \bea {\begin{eqnarray}}
\def \eea {\end{eqnarray}}
\def \ba {\begin{eqnarray*}}
\def \ea {\end{eqnarray*}}

\title{Density ratio model with data-adaptive basis function}
\author{Archer Gong Zhang}
\author{Jiahua Chen}
\affil{Department of Statistics, University of British Columbia}
\date{March 4, 2021}

\begin{document}
\maketitle

\begin{abstract} 
In many applications, we collect independent samples from interconnected populations. 
These population distributions share some latent structure, so it is
advantageous to jointly analyze the samples.
One effective way to connect the distributions is
the semiparametric density ratio model (DRM). 
A key ingredient in the DRM is that the log density
ratios are linear combinations of prespecified functions;
the vector formed by these functions is called the basis function.
A sensible basis function can often be chosen based on knowledge of the context, and
DRM-based inference is effective even if the basis function is imperfect.
However, a data-adaptive approach to the choice of basis function
remains an interesting and important research problem. 
We propose an approach based on the classical functional principal component analysis (FPCA). 
Under some conditions, we show that this approach leads to consistent basis function estimation.
Our simulation results show that the proposed adaptive choice leads to an efficiency gain.
We use a real-data example to demonstrate the efficiency gain and the ease of our approach.
\end{abstract}


%Main body of the paper 
\section{Introduction} 
\label{sec:intro}

This research is motivated by applications where
multiple samples are collected from connected or similar populations. 
For example, the income distribution of individuals or households is a key economic indicator 
for a country. 
Economists are keenly interested in changes over time in income distributions.  
One may postulate a parametric model for these distributions, but
a mild violation of the model assumptions may lead to dubious conclusions.
Nonparametric approaches avoid
such risks, at the cost of statistical efficiency. 
Semiparametric approaches offer a middle ground. 
\citet {chen2013quantile} propose connecting the population distributions via the 
density ratio model (DRM) of \citet {anderson1979multivariate}. 
Specifically, let the population distributions be $ G_{k} (\cdot) $ for population $ k = 0, \ldots, m $, 
and let their respective density functions with respect to some $ \sigma $-finite measure be $ g_{k} (\cdot) $. 
The DRM postulates that  
\begin{align}
g_{k} (x) = g_{0} (x) \exp \{ \alpha_{k} + \bbeta_{k}^{\top} \bq (x) \}, 
\hspace{5mm}
k = 1, \ldots, m,
\label{DRM}
\end{align}
for some known vector-valued basis function $ \bq (\cdot) $ and unknown parameters $ (\alpha_{k}, \bbeta_{k}) $. 
We refer to distribution $ G_0 (\cdot) $ as the base distribution, but any $ G_k (\cdot) $
could serve the same purpose.
The DRM is closely connected with other semiparametric models such as 
the biased sampling model \citep {vardi1982nonparametric, vardi1985empirical, qin2017connections},
the exponential tilting model \citep {rathouz2009generalized},
and the proportional likelihood ratio model \citep {luo2012proportional}.
The DRM characterizes the common features of the multiple populations through the
basis function $ \bq (\cdot) $. 
There have been many interesting developments in inference under the DRM 
\citep {qin1993empirical, qin1997goodness, fokianos2001semiparametric, 
qin1998inferences, chen2013quantile, cai2017hypothesis, de2017bayesian, sugiyama2012density}.
The model has been adopted in a broad range of applications, including
time series \citep {kedem2008forecasting},
finite mixture models \citep {tan2009note, liliuqin2017semiparametric},
mean regression models \citep {huang2012proportional},
the detection of changes in lower percentiles \citep {chen2016monitoring},
and the study of dominance relationships among multiple populations \citep {zhuang2019semiparametric}.

There has been much interest in both theory and methodological developments, and
one challenging research problem is the choice of $ \bq (\cdot) $.
The existing results are usually obtained under the assumption that $ \bq (\cdot) $ is correctly specified.
There are also natural choices that lead to reliable inference conclusions.
For example, one may set $ \bq (x) = (x, x^{2})^{\top} $ when the data histograms
resemble those of normal densities or use $ \bq (x) = (|x|^{1/2}, x, x^{2}, \log (1+|x|))^{\top} $ to cover a large territory.
\citet {fokianos2007density} proposes choosing the basis function via selection criteria, 
such as the Akaike information criterion \citep[AIC;][]{akaike1973information} 
or the Bayesian information criterion \citep[BIC;][]{schwarz1978estimating}. 
\citet {chen2013quantile} suggest that if the basis function is rich but not perfect,
empirical-likelihood-based quantile estimators are still more efficient than the sample quantiles. 
However, in some applications model misspecification is still an issue.
\citet {fokianos2006effect} and \citet {fokianos2007density} report increased bias in the estimation of model parameters 
and power loss in hypothesis tests if the basis function is underfitted, overfitted, or misspecified.
An adaptive basis function can be useful in such situations.

In this paper, we propose an approach based on functional principal component analysis 
\citep[FPCA;][]{ramsay2005functional, wang2016functional}.
FPCA identifies variability and similarity among the functions of interest through an ``optimal'' expansion of these functions.
Accurate approximations can be obtained by linear combinations of a few eigenfunctions.
For the DRM, we identify the most important eigenfunctions
of the log density ratios $ \{ \log (g_{k} (x)/g_{0} (x)): k = 0, \ldots, m \} $ through multiple samples
and thus obtain a data-adaptive $ \bq (\cdot) $.
Under the assumption that the log density ratios belong to a low-dimensional space,
the proposed estimators are consistent.
Our simulation experiments show that when the data are generated from
commonly used distributions, DRM analysis based
on an adaptive basis function has a promising efficiency.
In fact, it is often comparable with an analysis based on the ``most suitable'' basis function. 
Naturally, if the natural basis function for a 
multiple-sample dataset is unsuitable, the adaptive choice
has a superior performance. Our real-data example confirms this finding.

%%%%%%%%%%%%%%%%%%%%%%%%%%%%%%%%%%%%%%%%%%%%%%%%%%%%%%%%%%%%%%%%%%%%%%

\section{Basis functions for DRM and FPCA} 
\label{sec:DRM_BasisFun}

\subsection{Overview of DRM}

Let $ \{ x_{k j}: j = 1, \ldots, n_{k}; k = 0, \ldots, m \} $ be $ m+1 $ independent samples, 
each consisting of independent and identically distributed (i.i.d.) observations from a distribution $ G_{k} $. 
We assume that the population distributions $G_k$ satisfy the DRM as given in \eqref{DRM}.
However, we do not assume knowledge of the basis function $ \bq (\cdot) $.
Let $ N = n_0 + \cdots + n_m $.
We propose a data-adaptive choice of $ \bq (\cdot) $ under the assumption that
$ \rho_k = n_k /N $ has a nonzero limit for each $ k $ as $ N \to \infty $,
while $ m $ is constant. Without loss of generality, we will regard $ \rho_k $
as a constant not evolving with $ N $. This is sensible because the asymptotic results 
apply when $ N $ is large and the proportions $ \rho_k $ are
not too close to zero. How the inference evolves with $ N $ is not an issue in applications.
With this understanding, we define 
\[
\bar G (x) = \sum_{k=0}^m \rho_k G_k(x)
\]
 to be a mixture distribution by regarding the $ G_k (\cdot) $ as subpopulation distributions.
Let $ X $ be a generic random variable. We denote
\[
\bar E \{ h(X)\} = \int h(x) \mathrm {d} \bar G (x)
\]
whenever the integration is well defined, for a generic function $h(\cdot)$.
Next, we assume that the unknown vector-valued basis function satisfies
$
\bar E \{ \bq(X) \} = \bsm {0}.
$
If it does not, we can write
\[
\alpha_k + \bbeta ^\top \bq(x) 
= \alpha_k + \bbeta^\top \bar E \{ \bq(X)\} 
+ \bbeta^\top \big [ \bq (x) - \bar E \{ \bq(X) \} \big ]
\]
and replace $ \bq (\cdot) $ by $ \bq (\cdot) - \bar E \{ \bq(X) \} $ to ensure that the assumption is satisfied when  the expectation is finite.

 After we obtain a data-adaptive choice of $ \bq (\cdot) $, we will reuse the data to
 fit the DRM \eqref {DRM} to estimate $ (\alpha_k, \bbeta_k) $ based on
 the empirical likelihood; see $\S$\,\ref{subsec:EL_DRM}.
 We first need some FPCA theory.

\subsection{Functional principal component analysis} 
\label{subsec:FPCA}

Let $ T $ be a closed subset of $ \mathbb {R} $,
$ \mathscr {B} (T) $ the Borel $ \sigma $-algebra on $ T $, 
and $ \Mcal $ a probability measure, 
such that $ (T, \mathscr {B} (T), \Mcal) $ forms a measure space.
Let $ L^{2} (T, \Mcal) $ be the space of all measurable real-valued
functions on $T$ such that
$
\int_{T} f (t) \Mcal (\mathrm {d} t) = 0
$ 
and 
$
\int_{T} f^{2} (t) \Mcal (\mathrm {d} t) < \infty.
$
With $\Mcal$ a probability measure, all square integrable
functions can be centralized. Hence, we have not excluded many
functions from the usual $ L^{2} (T, \Mcal) $ space.
For any $ f, g \in L^{2} (T, \Mcal) $, define the inner product to be 
\[
\langle f, g \rangle = \int_{T} f (t) g (t) \Mcal (\mathrm {d} t).
\]
According to the comment after Definition 1.6 of \citet{conway2007course}, 
$ L^{2} (T, \Mcal) $ is a Hilbert space with norm $ \| f \|_{2} = (\langle f, f \rangle)^{1/2} $.

Let $ \Sigma (s, t) $ be a symmetric, nonnegative definite, and square integrable 
function with respect to $ \Mcal $, 
such that 
$
\int_{T} \int_{T} \Sigma^{2} (s, t) \Mcal (\mathrm {d} s) \Mcal (\mathrm {d} t) < \infty,
$
and for any $ s, t \in T $, 
$ 
\int \Sigma (s, t)  \Mcal (\mathrm {d} s) =  \int \Sigma (s, t) \Mcal (\mathrm {d} t) = 0.
$
For instance, $\Sigma (s, t) = (s- c)(t-c)$ with $c = \int t \Mcal (\mathrm {d} t) $ is such a
bivariate function when $\Mcal$ has a finite second moment. 
Next, we define a Hilbert--Schmidt integral operator $\Vcal$
\citep {arveson2006short} as
\begin{align} 
[\mathcal {V} f] (t) = \int_{T} \Sigma (s, t) f (s) \Mcal (\mathrm {d} s), 
\label {cov_opt}
\end{align} 
for any function $ f \in L^{2} (T, \Mcal) $. 
It can be seen that $[\mathcal {V} f] (t)  \in  L^{2} (T, \Mcal) $.

In the context of the DRM, 
let $ q_1 (\cdot), \ldots, q_d (\cdot) $ be $ d $ distinct and orthonormal functions in $ L^{2} (T, \Mcal) $.
Let
$
\Sigma (s, t) = \sum_{j = 1}^{d} \lambda_{j} q_{j} (s) q_j (t)
$
for some positive values $ \lambda_1, \ldots, \lambda_d $.
Clearly, its Hilbert--Schmidt integral operator $ \Vcal $
has $ \{ \lambda_j, q_j (t) \}_{j=1}^d $ as its eigenvalues and eigenfunctions.
Further, $ [\mathcal {V} f] (t) $ is always a linear combination of $ \{ q_j (t) \}_{j=1}^d $. 
Hence, $ \Vcal $ has no other nonzero eigenvalues.
By the spectral theorem, any $ f \in L^{2} (T, \Mcal)$ can be decomposed as 
$
f (t) = \sum_{j=1}^d a_j q_j (t) + h (t),
$
for some $ h (t) $
such that $ [\mathcal {V} h] (t) =0 $, and $ a_j = \langle f, q_j \rangle $.

\subsection{Applying FPCA to DRM}

We now revise the DRM assumption so that the meaning of the notation in \eqref{DRM}
aligns with that in $\S$\,\ref{subsec:FPCA}. 
Let $ \Xcal $ be the common support of population distributions $ \{ G_k (\cdot): k = 0, \ldots, m \} $.
We assume that there exist $ d < m $ distinct and orthonormal functions $ q_1 (\cdot), \ldots, q_d (\cdot) $
of $ L^{2} (\Xcal, \bar{G}) $ and unknown parameters
in the form $ (\alpha_k, \bbeta_k) $ such that
\[
Q_k (x) = \log \{ g_k (x)/g_0 (x) \} = \alpha_k  + \bbeta_k^\top \bq(x),
\]
for $ k = 0, \ldots, m $, where $ (\alpha_{0}, \bbeta_{0}) = \bsm {0} $ and $ \bq (x) = (q_{1} (x), \ldots, q_{d} (x))^{\top} $. 

By the Gram--Schmidt orthogonalization process \citep {conway2007course}, 
the existence of $ \bq (\cdot) $ with these properties is guaranteed.
We have required only that $ \bq (\cdot) $ is square integrable with respect to
$ \bar G $. This condition imposes a minimum restriction on the DRM:
the second moment of $ \bq (X) $ must be finite.
We hope that $ d < m $ so that the DRM itself is meaningful
in applications. 
For conceptual convenience and notational simplicity, denote
\[
(\bar \alpha, \bar \bbeta) =  (m+1)^{-1} \sum_{k=0}^m (\alpha_k, \bbeta_k);
\hspace {5mm}
\bar Q (x) = (m+1)^{-1} \sum_{k=0}^m Q_k (x) = \bar \alpha + \bar \bbeta^\top \bq(x),
\]
and introduce
\bea
\label{QQ}
Q^c_k (x) = \{ Q_k (x) - \alpha_k \} -  \{\bar Q (x) - \bar \alpha \}
=  \{ \bbeta_k - \bar \bbeta \} ^\top \bq(x).
\eea
The additional centralization step in the above definition is needed to
ensure the symmetry of $ G_0, \ldots, G_m $.
That is, the inference will not be affected by which population is regarded as $ G_0 $.
We assume that the rank of the linear space formed by 
$\{ Q^c_0 (x), \ldots, Q^c_m (x)\}$
is $d$; if not, we simply replace $\bq(x)$ by one of 
the orthonormal bases of this space.

We now connect these definitions with FPCA.
Let 
\begin{equation}
\label{Sigma}
\Sigma (s, t) 
= \sum_{k=0}^m Q^c_k (s) Q^c_k(t).
\end{equation}
Under the finite second moment condition,
we have
$
\int_\Xcal \int_\Xcal
\Sigma^{2} (s, t) \mathrm {d} \bar G (s) \mathrm {d} \bar G (t) < \infty.
$
(From now on, we will not specify the range of integration if it is obvious.)
This leads to the corresponding Hilbert--Schmidt integral operator $\Vcal$
defined in \eqref{cov_opt}.
For any $ f \in  L^{2} (\Xcal, \bar{G}) $, it can be seen that
\begin{align}
[ \Vcal f ] (t)
=
 \int \big \{ \sum_{k=0}^m Q^c_k (s) Q^c_k (t) \big \} f (s) \mathrm {d} \bar G (s) 
=
\sum_{k=0}^m \big \{ \int Q^c_k (s) f (s) \mathrm {d} \bar G (s) \big \} Q^c_k (t),
\label {int_operator}
\end{align}
which is a linear combination of the functions $ \{ Q^c_k (t): k = 0, \ldots, m \} $.
By definition of $ Q^c_k (\cdot) $, the output of the operator
$ [ \Vcal f ] $ is a linear combination of 
the functions in $ \bq (\cdot) $. This implies that all the eigenfunctions of $ \Vcal $
with nonzero eigenvalues are in the linear space formed by $ \bq (\cdot) $. 
It is clear that the rank of the linear space of eigenfunctions with 
nonzero eigenvalues is $ d $.

In the problem of interest, we do not have full knowledge of
$ \Sigma (s, t) $ defined by \eqref{Sigma}. However, for every $k$ we can
consistently estimate $ Q^c_k (\cdot) $  based
on data and thus obtain a consistent estimate of $ \Sigma (s, t) $.
We can then find an orthonormal basis of the nonzero 
eigenspace of $ \Vcal $, namely $ \bq (\cdot) $.

Before we discuss estimation problems, we explain how to recover $ \bq (\cdot) $ if we are given complete knowledge
of $ \{ Q^c_k (\cdot): k = 0, \ldots, m \} $. 
Let $ \bM $ be an $ (m+1) \times (m+1) $ matrix with $(i, j)$th element 
\bea
\label{M_matrix}
\bM(i, j) = \langle Q^c_i , Q^c_j \rangle = \bar E \{ Q^c_i(X) Q^c_j(X)\},
\eea
where $ i, j $ range from $ 0 $ to $ m $.
By the conditions stated following \eqref{QQ}, the rank of $\bM$ is $d$.
Let 
$ 
\{ \bp_{j} = (p_{0, j}, \ldots, p_{m, j})^\top : j = 0, \ldots, d-1 \} 
$
be a set of orthonormal eigenvectors of
$\bM$ corresponding to the eigenvalues
$\lambda_0 \geq \cdots \geq \lambda_{d-1}$.
The following theorem reveals the connection between the eigensystem of $ \bM $ and the DRM basis function $ \bq (x) $.
\begin{theorem}
\label{thm.MV}
Given the notation specified in this section and the DRM assumptions, we have

\textup {(i)} the nonzero eigenvalues of $\Vcal$ are also $ \lambda_0 \geq \cdots \geq \lambda_{d-1} $;

\textup {(ii)} for $ j = 0, \ldots, d-1 $, the functions
$
\psi_j (x) = \lambda_j^{-1/2} \sum_{k=0}^m p_{k, j} Q_k^c (x)
$
form a set of orthogonal, norm-one eigenfunctions of $ \Vcal $ with nonzero eigenvalues;

\textup {(iii)} every function in $ \bq (x) $ is a linear combination of $ \{ \psi_j (x): j = 0, \ldots, d-1 \} $.
\end{theorem}

Theorem~\ref{thm.MV} motivates a natural
way to construct an adaptive basis function for statistical inference under the DRM;
the proof is in the Appendix.

\section{Adaptive basis function} 
\label{sec:Est_Eigen}

\subsection{Estimation of eigenfunctions}

Following the developments in $\S$\,\ref {sec:DRM_BasisFun}, we
first consistently estimate the eigenfunctions of $ \Vcal $
with nonzero eigenvalues. 
These will serve as an adaptive
basis function for data analysis based on the DRM.
There are numerous statistical, mathematical, and
numerical issues. We first address the issue of
consistent estimation of eigenfunctions.

Recall that $ \{ x_{r j}: j = 1, \ldots, n_{r} \} $ is a random sample from population $ G_r $.
We estimate the density function via the kernel method \citep {rosenblatt1956, parzen1962estimation}:
\begin{align}
\label{kernel+}
\hat g_{r} (x) 
= \max \Big \{
\frac {1} {n_{r} h_{r}} \sum_{j = 1}^{n_{r}} K \big ( \frac {x - x_{r j}} {h_{r}} \big ), 
C_{r} (\log N/N)^{2/5}
\Big \},
\end{align}
for some kernel function $ K (\cdot) $, positive constant $ C_{r} $, and kernel bandwidth $ h_{r} $.
We modify the usual kernel density estimator slightly to avoid its value 
at any point being too close to $ 0 $ for the consistency proof needed later.
This technical step limits the value of $ \log \hat g_r (x)$ even though
$\hat g_r (x)$ is no longer a density function after this. 
It has purely technical significance, and we do not implement
this step in the data analysis.
Under mild conditions on $ g_r (x) $ and for appropriate
choices of the kernel function and bandwidth, 
the kernel density estimator $ \hat g_{r} (x) $ is consistent.
The specific conditions will be given in $\S$\,\ref {sec:asymptotics}.

Let 
$
\bar F_{n} (x) = N^{-1} \sum_{k, j} \mathbbm {1} (x_{k j} \leq x) 
$
be the empirical distribution of the pooled data 
$ \{ x_{k j}: j = 1, \ldots, n_{k}; k = 0, \ldots, m \} $. 
We now estimate the plain log density ratio $Q_k(x)$ by
\[ 
\hat Q_{k} (x) 
= 
\log \{ \hat g_{r} (x)/\hat g_{0} (x) \} 
- 
\int \log \{ \hat g_{r} (x)/\hat g_{0} (x) \} 
 \mathrm {d} \bar F_{n} (x)
\]  
and its centralized version by
\[ 
\hat {Q}^c_k (x) =\hat Q_{k} (x) - (m+1)^{-1} \sum_{k=0}^m \hat Q_{k} (x).
\]
We then estimate the $\bM$ matrix with $(i, j)$th element
\begin{align} 
\label {M_hat}
\hat {\bM} ({i, j})
=
\int \hat {Q}^c_i (x)  \hat {Q}^c_j (x) \mathrm {d} \bar F_{n} (x). 
\end{align}  
We can see that $ \hat \bM $ is an empirical 
version of $ \bM $ defined in \eqref {M_matrix}.  
Let the $d$ eigenvectors of $ \hat \bM $ be 
$
 \{ \hat \bp_{j} 
 = (\hat p_{0, j}, \ldots, \hat p_{m, j})^\top: j = 0, \ldots, d-1 \},
$
with eigenvalues 
$\hat \lambda_{0} \geq \cdots \geq \hat \lambda_{d-1}$.

We may define an empirical integral operator in the same spirit, but it does not seem to be needed. 
Conceptually, in view of Theorem \ref{thm.MV}, we introduce the estimated eigenfunctions
\begin{align}
\label {eigenfn_hat}
\hat \psi_{j} (x)
=
\hat \lambda_{j}^{-1/2} \sum_{k = 0}^{m} \hat p_{k, j} \hat Q^c_{k} (x),
\end{align}
for $ j = 0, \ldots, m $. 
If the estimators are consistent, we should have $\hat \lambda_{d-1} > 0$.
In a real-data analysis, we may have one or more $\hat \lambda_{d-j}$
very close to $ 0 $ by some standard.
If so, we could reduce the dimension of the
basis function in the DRM assumption and reduce the assumed $d$
in the subsequent estimation.

Given $ d $, we recommend a data-adaptive vector-valued basis function
\begin{align}
\label{adapt.basis}
\hat \bq(x) = (\hat \psi_{0} (x), \ldots, \hat \psi_{d-1} (x))^\top.
\end{align}
In this paper, we assume knowledge of $d$ in the theoretical development.
In $\S$\,\ref{subsec:adaptive_nFPC} we suggest some data-adaptive approaches to choosing $ d $.

\subsection{Interpretation of estimated eigenfunctions}

We trust that the variations between the log density ratios $ \{ Q_{r}^{c} (\cdot): r = 0, \ldots, m \} $ can be
well explained by the eigenfunctions corresponding to the
largest eigenvalues, namely the functional principal components.
Otherwise, the DRM will not be effective and is not recommended.
In the data analysis, $ \hat \bM $ likely has $ m $ nonzero eigenvalues.
By the theory of FPCA, the eigenfunctions corresponding to the top 
$d$ eigenvalues explain the largest proportion of variation among 
all the function spaces of dimension $d$. 
Mathematically, the orthonormal basis formed by the eigenfunctions 
$ \{ \hat \psi_{j} (\cdot): j = 0, \ldots, d-1 \} $ minimizes the empirical mean squared error  
\[ 
\sum_{r = 0}^{m} \int
\Big [ \hat Q_{r}^{c} (x) - \sum_{j = 0}^{d-1} a_{r, j} f_{j} (x) \Big ]^{2} \mathrm {d} \bar F_{n} (x) 
\] 
with respect to all possible orthonormal bases $ \{ f_{j} (\cdot): j = 0, \ldots, d-1 \} $,
where $ a_{r, j} = \int \hat Q_{r}^{c} (x) f_{j} (x) \mathrm {d} \bar F_{n} (x) $.
Hence, FPCA resembles classical principal component analysis. 

\subsection{Adaptive choice of number of eigenfunctions}
\label{subsec:adaptive_nFPC}

The rank $d$ exists only in the theoretical development. If we regard the DRM
as a gentle expansion of the normal model, then we use $ \bq (x) = (x, x^2)^{\top} $ and $d=2$.
Otherwise, we must select a $d$ based on the data.
By the theory of FPCA, the ratio 
$ \hat \lambda_{j}/\sum_{j' = 0}^{m} \hat \lambda_{j'} $ 
is the proportion of variation explained by the $ j $th functional principal component.
We therefore suggest choosing $d$ to be the smallest $J$ such that
$ \sum_{j = 0}^{J-1} \hat \lambda_{j}/\sum_{j' = 0}^{m} \hat \lambda_{j'} $
exceeds a threshold such as $ 90\% $ or $ 95\% $.

Information criteria such as AIC and BIC are commonly used in these situations.
In $\S$\,\ref {sec:GoodnessOfFit}, we will develop a
profile log empirical likelihood $ \tilde \ell (\cdot) $ of the unknown model parameters $ (\balpha, \bbeta) $ 
for the purposes of inference.
Given a basis function $ \bq (x) $ of dimension $J$ and
with $ (\hat \balpha, \hat \bbeta) $ the maximum likelihood estimator,
we have a natural BIC for $ \bq (x) $:
\[
\mathrm {BIC} (\bq) = -2 \tilde \ell (\hat \balpha, \hat \bbeta) + m J \log N.
\]
BIC favours models that achieve a trade-off between complexity (measured by a cost of $ \log N $ 
per extra parameter)
and fitting the data well in terms of the likelihood.
Our second proposal is therefore to choose up to $J$ principal eigenfunctions that 
minimize $\mathrm {BIC} (\bq) $.

A good model fit has two components. 
We must try to catch the most variation between the log density ratios in the populations,
and we must ensure that the model fits  but does not overfit the data.
These two goals may conflict, and we aim to address them both.
In the simulations and the real-data example,
we use a 95\% threshold as discussed above to obtain $J_1$, and
we use BIC to obtain $J_2$ from $ \{ 1, 2, 3, 4 \} $. 
We then set $d = \max \{J_1, J_2\}$ for the data analysis. 
This worked well in our simulations and the real-data example.
If the plain BIC asks for $d > 4$, then
no meaningful latent structure is available.
One should then consider whether the DRM is an appropriate model.

\subsection{Adaptive choice of kernel bandwidth}
\label {subsec:adaptive_bw}

The kernel bandwidth in \eqref{kernel+} is an important factor in the data analysis. 
If density estimation is the ultimate goal, \citet{silverman1986density} 
recommends a well-known rule-of-thumb bandwidth, and this choice is optimal when the true distribution is normal.
It remains a good choice for a broad range of density functions.
Unfortunately,  based on pilot simulations, we find that this choice
leads to nonsmooth basis functions (or ones with a large total variation). 
The subsequent DRM-based inference has unstable performance.

We instead use an adaptive version of \citet{silverman1986density}.
Suppose we know the true eigenvalues and eigenfunctions. 
We may then choose a bandwidth such that the estimated eigensystem
closely matches the known eigensystem. Based on this idea, we suggest minimizing
\begin{align} 
\sum_{j = 0}^{d-1} 
\int [ {\hat \lambda_{j}}^{-1/2} \hat \psi_{j} (x) - {\lambda_{j}}^{-1/2} \psi_{j} (x)]^{2} \mathrm {d} \bar F_{n} (x).
\label {bw_min}
\end{align} 
We consider the samples in an application to be generated from either normal or gamma distributions,
based on background information.
Specifically, we calculate the log density ratios $ \{ Q_{r}^{c} (\cdot): r = 0, \ldots, m \} $
using normal or gamma density functions.
We then obtain the ``true'' eigensystem characterized by $ \{\lambda_j, \psi_j (\cdot): j = 0, 1 \} $.
Then, \eqref {bw_min} with $ d = 2 $ is well defined.
To reduce the computational cost, we set $ h_{r} = k n_{r}^{-1/5} \hat \sigma_{r} $ for some
$k$ to be decided, with $ \hat \sigma_{r} $ being the sample standard deviation of the $ r $th sample, 
for $ r = 0, \ldots, m $.
We then minimize \eqref {bw_min} with respect to a single variable $ k $.
In our simulations and the real-data example, we use this adaptive choice in \eqref{kernel+}. 
It performs better than our other choices, which we do not report.

%%%%%%%%%%%%%%%%%%%%%%%%%%%%%%%%%%%%%%%%%%%%%%%%%%%%%%%%%%%%%%%%%%%%%%

\section {Asymptotic properties of adaptive basis function}
\label{sec:asymptotics}

In this section, we show that, under some conditions, the space formed by $ \{ \hat \psi_{j} (x): j = 0, \ldots, d-1 \} $ 
defined in \eqref {eigenfn_hat} converges to the space formed by 
$ \{ \psi_{j} (x): j = 0, \ldots, d-1 \} $.
We present some asymptotic results here; the proofs are given in the Appendix.

\begin{lemma}
\label{lemma_kernel}
Suppose we have $m+1$ random samples from multiple
populations satisfying the conditions on the DRM given
earlier. We assume that the population density functions
$g_{0} (x), \ldots, g_{m} (x) $ have bounded second derivatives. 

Moreover, the kernel function $K(\cdot)$ and bandwidth $h_r$ must satisfy
the following conditions:

\textup {(A1)}
$ K (\cdot) $ is  symmetric, is bounded, has a continuous and bounded first derivative,
and satisfies $\int x ^2 K(x) dx < \infty$.

\textup {(A2)}
$ h_r = B_r (\log N/N)^{1/5} $ for some positive constant $ B_r $ for $ r = 0, \ldots, m $.

Then, as the total sample size $ N \to \infty $ and for $ r = 0, \ldots, m $, with the kernel
density estimator $ \hat g_{r} (x) $ defined by \eqref {kernel+}, we have 
$
\sup_x |\hat g_r (x) -  g_r (x) | = O \big ( (\log N/N)^{2/5} \big )
$
almost surely.
\end{lemma}

Density functions for which $ g_r (x) $ does not satisfy the lemma conditions 
exist in theory but are likely not of concern in applications. 
For most kernel functions,
the above lemma imposes specific conditions that are sufficient
to apply Theorem 2.1.8 of \citet[p. 48]{PrakasaRao1983}
with $s = 1$ and $\alpha = 1$ leading to $\beta = 2$.
The original result requires $K(x) = 0$ when $|x| > A$ for some
constant $A$. 
We intend to use the standard normal density as the
kernel function, so we wish to avoid this condition.
The book contains an easy to comprehend proof (pp.~47--48),
and we judge that the key inequalities remain valid when 
$ K (\cdot) $ also has a bounded first derivative and a finite second moment.
Another remark concerns the truncation operation in \eqref{kernel+}.
It does not change the uniform rate because this estimator differs from
the original kernel density estimator by no more than $ O ((\log N/N)^{2/5}) $.

Our next result establishes the consistency of $ \hat \bM $ defined in 
\eqref {M_hat} for $\bM$ defined in \eqref{M_matrix}. 

\begin{theorem}[Consistency of $ \hat \bM $] 
\label {consist_M}
Under the conditions in Lemma~\ref{lemma_kernel},
assume that the density functions $g_r(\cdot)$ have finite upper bounds and that
$
\bar E \big | \log g_r(X) \big |^3 < \infty
$
for $ r = 0, \ldots, m $.
Further, assume that the common support $\Xcal$ is an interval of real numbers,
and that there is a positive constant $C$ not depending on $N$ such that the ratio
of the bandwidths $h_i/h_j < C$ for all $0 \leq i, j \leq m$.
Then, as $ N \to \infty $, we have
$
 \hat \bM({i, j}) - \bM({i, j}) \to 0
$
in probability for all $0 \leq i, j \leq m$.
\end{theorem} 

Our focus is on the relevance of the conclusion and the ease of understanding;
we therefore do not make the theorem conditions as weak as possible. 
In applications, one never knows whether the population distributions truly satisfy these conditions,
but we trust that all conceivable application examples satisfy them. 

The consistency of the matrix estimator $ \hat \bM $ leads to consistency 
of the eigensystem associated with $ \hat \bM $.
For definitiveness, we assume that a set of orthonormal eigenvectors 
with descending eigenvalues of $\bM$ have been fixed.
As before, they are denoted $ \lambda_j, \bp_j $ for $ j = 0, \ldots, d-1 $.
This is applicable even when $\bM$ has equal nonzero eigenvalues.

\begin{theorem}[Consistency of the eigensystem of $ \hat \bM $] 
\label {consist_eigen}

Under the conditions of Theorem~\ref {consist_M}, 
there exists a set of $ d $ eigenvectors $ \hat \bp_{0}, \ldots, \hat \bp_{d-1} $ of $ \hat \bM $,
with eigenvalues 
$ \hat \lambda_0 \geq \cdots \geq \hat \lambda_{d-1} $,
such that for $ j = 0, \ldots, d-1 $,
we have 
\textup {(i)} $ \hat \lambda_{j} - \lambda_{j} \to 0 $; 
\textup {(ii)} $ \hat {\bsm {p}}_{j} - \bsm {p}_{j} \to 0 $; 
and \textup {(iii)} for all $ x \in \Xcal, \hat \psi_{j} (x) - \psi_{j} (x) \to 0 $
as $ N \to \infty $.
\end{theorem} 

The eigenfunctions may not be unique, but the space that they span is. 
We recommend using
$ \{ \hat \psi_{j} (x): j = 0, \ldots, d-1 \} $ to form an adaptive basis function
(vector) when fitting the DRM.
The above theorem implies that the log density ratios 
$ \{ Q_{r}^{c} (x): r = 0, \ldots, m \} $ can be well approximated by
linear functions of the recommended adaptive basis function.

%%%%%%%%%%%%%%%%%%%%%%%%%%%%%%%%%%%%%%%%%%%%%%%%%%%%%%%%%%%%%%%%%%%%%%

\section{Data analysis with adaptive basis function}
\label{sec:GoodnessOfFit}

\subsection{Inference framework of EL under DRM} 
\label {subsec:EL_DRM} 

The DRM was proposed by \citet {anderson1979multivariate},  
and its theory and application have attracted much attention:
see the references in the Introduction.
For completeness, we provide a short description of the inference procedures here.
We assume the DRM as given in \eqref{DRM}, assume knowledge of $ d $, and use
$
\bq (x) = \bpsi (x) = (\psi_0 (x), \ldots, \psi_{d-1} (x))^\top.
$
There are $m+1$ independent random samples 
$ \{ x_{k j}: j = 1, \ldots, n_{k}; k = 0, \ldots, m \} $ with
population distributions $ \{ G_k (x): k=0, \ldots, m \} $ and
total sample size $ N = n_0 + \cdots + n_m $.
The sample proportions $ \rho_k = n_k/N $ for $ k = 0, \ldots, m $
remain constant for asymptotic considerations.

Denote $ p_{k j} = \mathrm {d} G_{0} (x_{ k j }) = \Pr (X = x_{k j}; G_{0}) $,
where the notation is no longer the elements of the eigenvectors of $ \bM $.
Following the
principle of the empirical likelihood \citep[EL;][]{owen2001empirical},
the EL under the DRM is given by
\[ 
L_{n} (G_{0}, \ldots , G_{m}) 
= \prod_{k, j} \mathrm {d} G_{k} (x_{k j}) 
= 
\big \{ \prod_{k, j} p_{k j} \big \}
\times 
\exp \Big \{ \sum_{k, j} [\alpha_{k} + \bbeta_{k}^{\top} \bq (x_{k j})] \Big \}, 
\] 
with $ \alpha_{0} = 0, \bbeta_{0} = \bsm {0} $ by convention. 
Since the EL function $ L_{n} (G_{0}, \ldots , G_{m}) $ 
is also a function of the base distribution $ G_{0} $ and model parameters 
$ 
\balpha^{\top} = (\alpha_{1}, \ldots, \alpha_{m}),
\bbeta^{\top} = (\bbeta_{1}^{\top}, \ldots , \bbeta_{m}^{\top}),
$
we may write its logarithm as 
\[ 
\ell_{n} (\balpha, \bbeta, G_{0}) 
= \log L_{n} (G_{0}, \ldots , G_{m}) 
= \sum_{k, j} \log (p_{k j}) + \sum_{k ,j} [\alpha_{k} + \bbeta_{k}^{\top} \bq (x_{k j})].
\] 
In all cases, the summations or products with respect to $ k, j $ are taken over the full ranges 
$ j = 1, \ldots, n_{k}; k = 0, \ldots, m$. 

Inference on the population parameters is mostly 
based on the profile likelihood. 
The DRM assumptions and the fact that the $ G_{r} $ are population distributions 
imply that for $ r = 0, \ldots , m $,
\[ 
\int \exp \big \{ \alpha_{r} + \bbeta_{r}^{\top} \bq (x) \big \} \, \mathrm {d} G_{0} ( x ) = 1.
\] 
By confining the support of $ G_{0}, \ldots , G_{m} $ to the observed data, 
we obtain the constraints 
\begin{align} 
\sum_{k, j} p_{k j} \exp \big \{ \alpha_{r} + \bbeta_{r}^{\top} \bq (x_{k j}) \big \} = 1.  
\label {constr1}   
\end{align} 

The profile log-EL of the model parameters $ (\balpha, \bbeta) $
is defined to be the maximum of the log-EL over $ G_{0} $ 
under constraints \eqref {constr1}: 
\[ 
\tilde \ell_{n} (\balpha, \bbeta) 
= 
\sup_{G_{0}} 
\Big \{ 
\ell_{n} (\balpha, \bbeta, G_{0}): \sum_{k, j} p_{k j} \exp \big \{ \alpha_{r} + \bbeta_{r}^{\top} \bq (x_{k j}) \big \} = 1,
~~ r = 0, \ldots, m 
\Big \}.  
\] 

Let $ (\hat \balpha, \hat \bbeta) $ be the maximizer of the profile log-EL $ \tilde \ell_{n} (\balpha, \bbeta) $. 
We then have the fitted values of $ p_{k j} $
\begin{align} 
\label{fitted_prob}
\hat p_{k j} 
= \Big [ \sum_{r = 0}^{m} n_{r} \exp \big \{ \hat \alpha_{r} + \hat \bbeta_{r}^{\top} \bq (x_{k j}) \big \} \Big ]^{-1} 
\end{align} 
and the fitted distribution functions
\begin{align} 
\hat G_{r} (x) 
= \sum_{k, j} \hat p_{k j} \exp \big \{ \hat \alpha_{r} + \hat \bbeta_{r}^{\top} \bq (x_{k j}) \big \} 
\mathbbm {1} (x_{k j} \leq x),
\label {DRM_EL_fit_distbn}
\end{align} 
where $ \mathbbm {1} (\cdot)$ is the indicator function.

With these functions, we carry out inference on the population
parameters by regarding them as functions of the population distributions.
In this paper, we focus on population quantiles and density functions
based on the fitted DRM with our adaptive basis function.

\subsection{Density and quantile estimation}
\label {subsec:QuantileDensity_Est}

We now study the EL-DRM inference with the proposed adaptive basis function.
We use the notation of \eqref{fitted_prob}
and \eqref{DRM_EL_fit_distbn}, except that $\bpsi(\cdot)$ is replaced
by $\hat {\bpsi}(\cdot)$ given in Theorem~\ref{consist_eigen}.

Many researchers have investigated density estimation;  
\citet {silverman1978weak} and \citet {wied2012consistency} 
are two representative references. 
\citet {kneip2001inference} give a convincing example of the relevance of density estimation 
in economics as well as FPCA-based inference.
After obtaining $\hat G_r(\cdot)$ through the EL-DRM approach with the
adaptive basis function, we update the density estimation via
\begin{align} 
\hat g_{r}^{\drm} (x) 
= \frac{1}{h_r^{'}} \int K \big( \frac{x-y}{h_r^{'}} \big ) \mathrm {d} \hat G_{r} (y),
\label {DRM_KDE}
\end{align} 
with $ K (\cdot) $ and $ h_r^{'} $ being as in previous sections.
We use Silverman's bandwidth $ h_{r}^{'} = 0.9 n_{r}^{-1/5} \min \{ \hat \sigma_{r}, \hat {\mathrm {IQR}}/1.34 \} $ \citep{silverman1986density},
with $ \hat \sigma_{r}, \hat {\mathrm {IQR}} $ being the standard deviation and interquartile range of the fitted distribution $ \hat G_{r} $.
Note that the kernel bandwidth in \eqref {DRM_KDE} is different from that in $\S$\,\ref {sec:Est_Eigen} (particularly \eqref {kernel+}).
In our simulations and real-data analysis, we set $ K (\cdot) $ to the standard normal density. 

Quantiles are of importance in many applications. 
\citet {chen2013quantile} study inference on quantiles under the 
DRM with prespecified basis functions $ \bq (\cdot) $.
We show that the approach remains valid and effective with our adaptive basis function. 
We estimate the quantiles at level $ \tau_r \in (0, 1) $ via
\begin{align} 
\hat \xi_{r} = \inf \{ t: \hat G_{r} (t) \geq \tau_{r} \}. 
\label {DRM_EL_quan_est}
\end{align} 
We refer to $ \{ \hat \xi_{r}: r = 0, \ldots, m \} $ as the EL-DRM quantile estimators. 
When the populations satisfy the DRM assumption with a correctly specified $ \bq (\cdot) $,
these estimators are consistent and asymptotically normal \citep{chen2013quantile}.
They are more efficient (i.e., have smaller variances) than the empirical quantiles. 
With the adaptive basis function $ \hat \bq (\cdot) $, the corresponding estimators should remain consistent.
They remain competitive in terms of efficiency, as our simulation studies will show.

\subsection{Other estimators}
\label {subsec:other_est}

We regard the estimators of \citet {kneip2001inference} as a competitor
and give a brief overview here.
They also consider the situation where multiple samples are available,
drawn from distributions that have some common features.
Instead of the DRM, they assume that the density functions $ \{ g_{r} (\cdot): r = 0, \ldots, m \} $ 
have a low-dimensional representation:
\begin{align}
g_{r} (x) = (m+1)^{-1} \sum_{r = 0}^{m} g_{r} (x) + \sum_{j = 1}^{L} \theta_{r, j} \varphi_{j} (x),
\hspace {5mm}
r = 0, \ldots, m,
\label {KU}
\end{align}
for some $ L \leq m $, where $ \{ \varphi_{j} (\cdot): j = 1, \ldots, L \} $ are orthonormal 
and obtained using FPCA via the Karhunen--Lo\`{e}ve expansion \citep {rice1991estimating}.
They first obtain estimates of $ \varphi_{j} (\cdot) $ and $ \theta_{r, j} $ by applying FPCA 
to the kernel density estimators of $ \{ g_{r} (\cdot): r = 0, \ldots, m \} $.
They substitute these into \eqref {KU} to update the density estimates.
\citet {kneip2001inference} do not discuss quantile estimation, 
although such estimates are readily available.
Unlike the DRM, model \eqref {KU} of \citet {kneip2001inference} 
is not compatible with commonly used models such as the normal or gamma
distribution family.
Moreover, their density estimates may be negative, which limits their use
in certain applications.

We also compare our estimators with the naive kernel density estimators with the Gaussian kernel and Silverman's bandwidth for nonparametric density estimation,
and the empirical quantiles for nonparametric quantile estimation.

%%%%%%%%%%%%%%%%%%%%%%%%%%%%%%%%%%%%%%%%%%%%%%%%%%%%%%%%%%%%%%%%%%%%%%

\section{Simulation Studies} 
\label{sec:simulations}

In this section, we use simulations to study the performance of the
EL-DRM inference with the proposed adaptive basis function.
We explore density estimation and quantile estimation. 

To evaluate the performance of a generic density estimator $ \hat g_r (\cdot) $ of $ g_{r} (\cdot) $, 
we use the integrated mean squared error (IMSE), 
which is calculated based on the simulation repetitions:
\begin{align*} 
\mbox{IMSE} (\hat g_r)
=
N_{\mathrm {rep}}^{-1} \sum_{\ell = 1}^{N_{\mathrm {rep}} }
\Big \{ \int [ \hat g_{r}  (x) - g_{r} (x) ]^{2} \mathrm {d} x \Big \}_{\ell},
\end{align*} 
where $ N_{\mathrm {rep}} $ is the number of simulation repetitions
and $ \{ \int [ \hat g_{r}  (x) - g_{r} (x) ]^{2} \mathrm {d} x \}_{\ell} $ is the output from the $ \ell $th simulation repetition.
We use the plain mean squared error (MSE) to measure
the performance of a generic quantile estimator $ \hat \xi_{r} $ of $ \xi_{r} $;
it is calculated as
\begin{align*} 
\mathrm {MSE} (\hat \xi_{r}) 
=
N_{\mathrm {rep}}^{-1} \sum_{\ell = 1}^{N_{\mathrm {rep}} }
\big \{ \hat \xi_{r} - \xi_{r}\big \}^2_{\ell},
\end{align*} 
where $\{ \hat \xi_{r} - \xi_{r}\}_{\ell}$ is the output from the
$\ell$th simulation repetition.

We consider scenarios where the data are from $ m+1 = 6 $ distributions,
satisfying the DRM with a basis function with $d<m$.
In the experiments, we generate $ 1000 $ sets of samples of sizes 
$ n_{r} = 500, 1000 $. 
{\bf Scenario 1}: The samples are from normal distributions 
with the means $ (18, 18.5, 18.5, 17.5, 19, 18) $ and an equal variance $ 6 $,
so the most suitable basis function is $ \bq (x) = x $.
{\bf Scenario 2}: The samples are from normal distributions with the
means $ (18, 18.5, 18.5, 17.5, 18, 18) $ and variances $ (6, 6.5, 7, 6, 6.5, 6) $.
With unequal variances, the most suitable basis function is $ \bq (x) = (x, x^{2})^{\top} $.
{\bf Scenario 3}: The samples are from gamma distributions with the shape parameters 
$ (6, 6, 7, 7, 8, 8) $ and scale parameters $ (1.5, 1.4, 1.3, 1.2, 1.1, 1.0) $. 
The most suitable basis function is $ \bq (x) = (x, \log x)^{\top} $. 
{\bf Scenario 4}: 
Let $ \phi (\cdot) $ be the density function of the standard normal 
and $ (\xi_{1}, \xi_{2}) = (-0.6745, 0.6745) $ the $25$th and $75$th percentiles.
The samples are from distributions with the density functions 
\[
g_{j} (x) 
= \phi (x) \exp \left \{ \alpha_{j} + \beta_{1, j} \phi (x-\xi_{1}) + \beta_{2, j} \phi (x-\xi_{2}) \right \},  
\hspace {5mm} 
j = 0, \ldots, 5, 
\]
with  $ (\beta_{1, 0}, \ldots, \beta_{1, 5}) = (0, 2, 1, 0, -2, 3)$, and
$(\beta_{2, 0}, \ldots, \beta_{2, 5}) = (0, -3, 1, -2, 2, -1) $. 
The parameters $ (\alpha_{0}, \ldots, \alpha_{5}) $ are normalizing constants.
The means, standard deviations, and shapes of these distributions are similar;
their densities are plotted in the Appendix.  
They satisfy the DRM with the most suitable
basis function being $ \bq (x) = (\phi (x-\xi_{1}), \phi (x-\xi_{2}))^{\top} $. 
Unlike the other scenarios, these distributions do not fit into any popular families,
but the proposed DRM with the adaptive basis function can still be used.
We call this a self-designed DRM.

We obtain the IMSEs of the density estimators and the simulated MSEs 
of the $10$th, $30$th, $50$th, $70$th, and $90$th percentiles;
for the latter we report only the average MSEs across the $ m+1 $ distributions to save space.
The simulation results are given in Table~\ref {combined_density_quan}.
We also include boxplots of these two performance measures 
in the Appendix.
In Table~\ref {combined_density_quan}, we use
``$d$ FPCs'' for the DRM with the adaptive basis function with $d$ eigenfunctions;
``Adaptive'' for the DRM with the adaptive basis function and the adaptive number of eigenfunctions as described in Section~\ref {subsec:adaptive_nFPC}; 
``Rich'' for the DRM with the rich basis function $ \bq (x) = (|x|^{1/2}, x, x^{2}, \log (1+|x|))^{\top} $;
 ``Truth'' for the DRM with the most suitable $ \bq (x) $ (given in the scenario description);
``K\&U $d$'' for the method of \citet {kneip2001inference} using $d$ orthonormal functions, as described in $\S$\,\ref {subsec:other_est};
and ``NP'' for the nonparametric estimators.
To save space, we include only the ``$d$ FPCs'' and ``K\&U $d$'' methods with the best-performing $d$.
In fact, the best-performing $d$ is the length of the most suitable $ \bq (x) $
in all the scenarios considered.

Table~\ref {combined_density_quan} shows that the DRM estimates based on
the adaptive basis functions and the best-performing $d$ have very low IMSE and MSE values.
They are satisfactorily close to the DRM estimates based on the most suitable basis functions. 
If the number of eigenfunctions is also selected adaptively,
the adaptive DRM remains highly competitive except in scenario 1.
In all the scenarios, the proposed fully adaptive DRM estimators (``Adaptive'' in the table) 
outperform the naive nonparametric estimators
when averaged over the $ 6 $ populations and $ 5 $ quantiles:
for density estimation the IMSE is $ 7\% $--$ 44\% $ smaller,
and for quantile estimation the MSE is $ 7\% $--$ 28\% $ smaller.
They considerably outperform the estimators of \citet {kneip2001inference} in all but two cases
(where the differences are marginal). 
We also observe that although the DRM estimates based on
the rich basis function have high efficiency when the data are from the normal and gamma distribution
families, they do not perform well in scenario 4. 
In this case, the rich basis function is substantially different from the most suitable basis function.
We have performed additional simulations for data generated from distributions 
that do not fit into the DRM with an appropriate-sized basis function;
these results are presented in the Appendix.
When the distributions have common features, 
the fully adaptive DRM estimators have an efficiency gain over the nonparametric estimators 
in both the density and quantile estimation.

\begin{table}%[h!] 
\caption{Simulated IMSEs of density estimators and average MSEs of quantile estimators scaled by respective sample sizes}
\begin{adjustbox}{max width=\textwidth}
\begin{tabular}{lrrrrrrrrrrrrrrr}
\multicolumn{1}{l}{Method}      
& \multicolumn{7}{c}{IMSE of density estimators}      
&
& \multicolumn{6}{c}{Average MSE of quantile estimators}       \\ 
& $ G_0 $ & $ G_1 $ & $ G_2 $ & $ G_3 $ & $ G_4 $ & $ G_5 $ & avg.      
&
& $ 10\% $ & $ 30\% $ & $ 50\% $ & $ 70\% $ & $ 90\% $ & avg. \\ 
%& \\ 
\multicolumn{15}{c}{Scenario 1: Normal with equal variances; $n_{r} = 500 $} \\ 
1 FPC      &  
$ 0.15 $ & $ 0.15 $ & $ 0.15 $ & $ 0.20 $ & $ 0.20 $ & $ 0.15 $ & $ 0.17 $
& &  
$ 9.73 $ & $ 7.28 $ & $ 6.90 $ & $ 7.20 $ & $ 9.37 $ & $ 8.10 $     
\\           
Adaptive      &  
$ 0.25 $ & $ 0.24 $ & $ 0.25 $ & $ 0.25 $ & $ 0.25 $ & $ 0.25 $ & $ 0.25 $
& &  
$ 13.67 $ & $ 8.39 $ & $ 7.26 $ & $ 8.08 $ & $ 13.31 $ & $ 10.14 $        
\\        
Rich      &  
$ 0.21 $ & $ 0.20 $ & $ 0.21 $ & $ 0.22 $ & $ 0.21 $ & $ 0.21 $ & $ 0.21 $    
& &  
$ 12.64 $ & $ 7.87 $ & $ 7.25 $ & $ 7.56 $ & $ 12.35 $ & $ 9.54 $         
\\                
Truth      &  
$ 0.14 $ & $ 0.14 $ & $ 0.14 $ & $ 0.15 $ & $ 0.14 $ & $ 0.14 $ & $ 0.14 $ 
& &  
$ 8.19 $ & $ 6.62 $ & $ 6.43 $ & $ 6.55 $ & $ 7.99 $ & $ 7.16 $      
\\     
K\&U 1      &  
$ 0.20 $ & $ 0.20 $ & $ 0.20 $ & $ 0.24 $ & $ 0.23 $ & $ 0.20 $ & $ 0.21 $
& &  
$ 15.83 $ & $ 8.20 $ & $ 6.89 $ & $ 8.26 $ & $ 15.40 $ & $ 10.92 $          
\\      
NP      &  
$ 0.39 $ & $ 0.38 $ & $ 0.38 $ & $ 0.39 $ & $ 0.37 $ & $ 0.39 $ & $ 0.38 $ 
& &  
$ 17.22 $ & $ 10.54 $ & $ 9.10 $ & $ 10.21 $ & $ 16.87 $ & $ 12.79 $       
\\   
%& \\ 
\multicolumn{15}{c}{Scenario 1: Normal with equal variances; $ n_{r} = 1000 $} \\ 
1 FPC      &  
$ 0.18 $ & $ 0.17 $ & $ 0.17 $ & $ 0.24 $ & $ 0.23 $ & $ 0.16 $ & $ 0.19 $
& &  
$ 9.61 $ & $ 7.44 $ & $ 6.93 $ & $ 7.27 $ & $ 9.50 $ & $ 8.15 $     
\\           
Adaptive      &  
$ 0.25 $ & $ 0.24 $ & $ 0.23 $ & $ 0.27 $ & $ 0.27 $ & $ 0.23 $ & $ 0.25 $
& &  
$ 12.19 $ & $ 7.95 $ & $ 7.03 $ & $ 7.64 $ & $ 11.78 $ & $ 9.32 $        
\\         
Rich      &  
$ 0.23 $ & $ 0.22 $ & $ 0.22 $ & $ 0.24 $ & $ 0.24 $ & $ 0.22 $ & $ 0.23 $        
& &  
$ 12.45 $ & $ 7.99 $ & $ 7.20 $ & $ 7.57 $ & $ 12.61 $ & $ 9.56 $         
\\               
Truth      &  
$ 0.17 $ & $ 0.16 $ & $ 0.16 $ & $ 0.17 $ & $ 0.17 $ & $ 0.16 $ & $ 0.17 $   
& &  
$ 8.10 $ & $ 6.74 $ & $ 6.49 $ & $ 6.64 $ & $ 8.01 $ & $ 7.20 $       
\\
K\&U 1      &  
$ 0.25 $ & $ 0.24 $ & $ 0.24 $ & $ 0.27 $ & $ 0.27 $ & $ 0.24 $ & $ 0.25 $       
& &  
$ 18.51 $ & $ 8.70 $ & $ 6.86 $ & $ 8.71 $ & $ 18.28 $ & $ 12.21 $               
\\     
NP      &  
$ 0.46 $ & $ 0.45 $ & $ 0.44 $ & $ 0.46 $ & $ 0.47 $ & $ 0.45 $ & $ 0.45 $   
& &  
$ 17.92 $ & $ 10.27 $ & $ 9.05 $ & $ 10.12 $ & $ 17.67 $ & $ 13.01 $        
\\   
%& \\ 
\multicolumn{15}{c}{Scenario 2: Normal with unequal variances; $ n_{r} = 500 $} \\       
2 FPCs      &  
$ 0.26 $ & $ 0.19 $ & $ 0.20 $ & $ 0.24 $ & $ 0.22 $ & $ 0.21 $ & $ 0.22 $            
& &  
$ 14.06 $ & $ 8.47 $ & $ 7.26 $ & $ 7.80 $ & $ 12.92 $ & $ 10.10 $         
\\    
Adaptive      &  
$ 0.29 $ & $ 0.21 $ & $ 0.21 $ & $ 0.25 $ & $ 0.23 $ & $ 0.23 $ & $ 0.24 $
& &  
$ 14.62 $ & $ 8.79 $ & $ 7.43 $ & $ 7.91 $ & $ 13.25 $ & $ 10.40 $      
\\            
Rich      &  
$ 0.22 $ & $ 0.20 $ & $ 0.20 $ & $ 0.22 $ & $ 0.20 $ & $ 0.21 $ & $ 0.21 $   
& &  
$ 12.74 $ & $ 8.02 $ & $ 7.35 $ & $ 7.61 $ & $ 12.62 $ & $ 9.67 $      
\\                     
Truth      &  
$ 0.20 $ & $ 0.17 $ & $ 0.17 $ & $ 0.19 $ & $ 0.17 $ & $ 0.18 $ & $ 0.18 $
& &  
$ 12.26 $ & $ 7.45 $ & $ 6.58 $ & $ 7.27 $ & $ 12.08 $ & $ 9.13 $      
\\
K\&U 2     &  
$ 0.41 $ & $ 0.24 $ & $ 0.26 $ & $ 0.34 $ & $ 0.29 $ & $ 0.25 $ & $ 0.30 $
& &  
$ 19.32 $ & $ 9.45 $ & $ 7.41 $ & $ 9.13 $ & $ 18.09 $ & $ 12.68 $     
\\       
NP      &  
$ 0.42 $ & $ 0.37 $ & $ 0.35 $ & $ 0.40 $ & $ 0.36 $ & $ 0.39 $ & $ 0.38 $
& &  
$ 17.43 $ & $ 10.68 $ & $ 9.24 $ & $ 10.35 $ & $ 17.09 $ & $ 12.96 $        
\\   
%& \\ 
\multicolumn{15}{c}{Scenario 2: Normal with unequal variances; $ n_{r} = 1000 $} \\      
2 FPCs      &  
$ 0.28 $ & $ 0.21 $ & $ 0.21 $ & $ 0.27 $ & $ 0.25 $ & $ 0.21 $ & $ 0.24 $            
& &  
$ 13.72 $ & $ 8.27 $ & $ 7.12 $ & $ 7.79 $ & $ 12.97 $ & $ 9.97 $         
\\    
Adaptive      &  
$ 0.31 $ & $ 0.21 $ & $ 0.21 $ & $ 0.27 $ & $ 0.26 $ & $ 0.22 $ & $ 0.25 $
& &  
$ 14.51 $ & $ 8.51 $ & $ 7.21 $ & $ 7.85 $ & $ 13.28 $ & $ 10.27 $       
\\                 
Rich      &  
$ 0.25 $ & $ 0.22 $ & $ 0.21 $ & $ 0.25 $ & $ 0.24 $ & $ 0.22 $ & $ 0.23 $      
& &  
$ 12.72 $ & $ 8.08 $ & $ 7.26 $ & $ 7.64 $ & $ 12.79 $ & $ 9.70 $        
\\                
Truth      &  
$ 0.22 $ & $ 0.19 $ & $ 0.18 $ & $ 0.22 $ & $ 0.20 $ & $ 0.19 $ & $ 0.20 $ 
& &  
$ 12.34 $ & $ 7.56 $ & $ 6.59 $ & $ 7.39 $ & $ 12.26 $ & $ 9.23 $      
\\
K\&U 2       &  
$ 0.49 $ & $ 0.25 $ & $ 0.27 $ & $ 0.41 $ & $ 0.34 $ & $ 0.24 $ & $ 0.34 $      
& &  
$ 20.87 $ & $ 9.54 $ & $ 7.29 $ & $ 9.30 $ & $ 19.66 $ & $ 13.33 $   
\\       
NP      &  
$ 0.50 $ & $ 0.43 $ & $ 0.41 $ & $ 0.48 $ & $ 0.45 $ & $ 0.45 $ & $ 0.45 $ 
& &  
$ 18.18 $ & $ 10.39 $ & $ 9.18 $ & $ 10.24 $ & $ 17.85 $ & $ 13.17 $        
\\   
%& \\ 
\multicolumn{15}{c}{Scenario 3: Gamma; $ n_{r} = 500 $} \\       
2 FPCs      &  
$ 0.16 $ & $ 0.20 $ & $ 0.20 $ & $ 0.19 $ & $ 0.26 $ & $ 0.22 $ & $ 0.20 $
& &  
$ 11.96 $ & $ 12.15 $ & $ 14.42 $ & $ 20.72 $ & $ 46.68 $ & $ 21.19 $
\\    
Adaptive      &  
$ 0.17 $ & $ 0.22 $ & $ 0.21 $ & $ 0.20 $ & $ 0.27 $ & $ 0.23 $ & $ 0.22 $
& &  
$ 12.37 $ & $ 12.64 $ & $ 14.90 $ & $ 20.99 $ & $ 46.32 $ & $ 21.44 $
\\             
Rich      &  
$ 0.18 $ & $ 0.21 $ & $ 0.19 $ & $ 0.20 $ & $ 0.20 $ & $ 0.20 $ & $ 0.20 $
& &  
$ 10.49 $ & $ 10.89 $ & $ 13.18 $ & $ 18.51 $ & $ 43.11 $ & $ 19.24 $
\\                                
Truth      &  
$ 0.14 $ & $ 0.15 $ & $ 0.14 $ & $ 0.15 $ & $ 0.16 $ & $ 0.17 $ & $ 0.15 $
& &  
$ 9.72 $ & $ 9.58 $ & $ 11.41 $ & $ 16.99 $ & $ 38.41 $ & $ 17.22 $
\\
K\&U 2      &  
$ 0.25 $ & $ 0.25 $ & $ 0.24 $ & $ 0.21 $ & $ 0.28 $ & $ 0.32 $ & $ 0.26 $
& &  
$ 21.85 $ & $ 11.21 $ & $ 13.59 $ & $ 22.53 $ & $ 52.95 $ & $ 24.43 $
\\       
NP      &  
$ 0.30 $ & $ 0.32 $ & $ 0.30 $ & $ 0.33 $ & $ 0.34 $ & $ 0.36 $ & $ 0.32 $
& &  
$ 14.26 $ & $ 13.51 $ & $ 16.30 $ & $ 23.30 $ & $ 53.36 $ & $ 24.14 $
\\   
%& \\ 
\multicolumn{15}{c}{Scenario 3: Gamma; $ n_{r} = 1000 $} \\          
2 FPCs      &  
$ 0.17 $ & $ 0.21 $ & $ 0.23 $ & $ 0.21 $ & $ 0.29 $ & $ 0.25 $ & $ 0.22 $           
& &  
$ 12.38 $ & $ 12.60 $ & $ 14.63 $ & $ 21.21 $ & $ 48.62 $ & $ 21.89 $          
\\    
Adaptive      &  
$ 0.18 $ & $ 0.22 $ & $ 0.23 $ & $ 0.22 $ & $ 0.30 $ & $ 0.25 $ & $ 0.23 $
& &  
$ 12.40 $ & $ 12.76 $ & $ 14.88 $ & $ 21.40 $ & $ 48.05 $ & $ 21.90 $       
\\      
Rich      &  
$ 0.20 $ & $ 0.21 $ & $ 0.21 $ & $ 0.22 $ & $ 0.22 $ & $ 0.24 $ & $ 0.22 $      
& &  
$ 10.40 $ & $ 11.15 $ & $ 13.22 $ & $ 18.66 $ & $ 43.35 $ & $ 19.35 $       
\\                                       
Truth      &  
$ 0.16 $ & $ 0.16 $ & $ 0.16 $ & $ 0.17 $ & $ 0.17 $ & $ 0.19 $ & $ 0.17 $  
& &  
$ 9.74 $ & $ 9.74 $ & $ 11.43 $ & $ 17.16 $ & $ 38.35 $ & $ 17.29 $       
\\
K\&U 2      &  
$ 0.29 $ & $ 0.28 $ & $ 0.28 $ & $ 0.23 $ & $ 0.30 $ & $ 0.39 $ & $ 0.29 $      
& &  
$ 24.12 $ & $ 11.42 $ & $ 13.61 $ & $ 22.92 $ & $ 57.31 $ & $ 25.88 $   
\\       
NP      &  
$ 0.35 $ & $ 0.37 $ & $ 0.37 $ & $ 0.40 $ & $ 0.39 $ & $ 0.44 $ & $ 0.39 $ 
& &  
$ 14.14 $ & $ 13.67 $ & $ 16.42 $ & $ 23.79 $ & $ 54.80 $ & $ 24.56 $       
\\   
%& \\ 
\multicolumn{15}{c}{Scenario 4: Self-designed; $ n_{r} = 500 $} \\        
2 FPCs      &  
$ 0.67 $ & $ 1.08 $ & $ 1.01 $ & $ 0.76 $ & $ 1.18 $ & $ 1.17 $ & $ 0.98 $
& &  
$ 2.62 $ & $ 1.45 $ & $ 1.30 $ & $ 1.60 $ & $ 3.39 $ & $ 2.07 $
\\    
Adaptive      &  
$ 0.79 $ & $ 1.16 $ & $ 1.04 $ & $ 0.75 $ & $ 1.18 $ & $ 1.21 $ & $ 1.02 $
& &  
$ 2.52 $ & $ 1.32 $ & $ 1.18 $ & $ 1.50 $ & $ 3.35 $ & $ 1.97 $
\\                                                 
Rich      &  
$ 0.98 $ & $ 4.05 $ & $ 1.00 $ & $ 0.91 $ & $ 6.03 $ & $ 2.71 $ & $ 2.61 $
& &  
$ 2.93 $ & $ 2.91 $ & $ 2.35 $ & $ 3.10 $ & $ 3.58 $ & $ 2.98 $
\\
Truth      &  
$ 0.58 $ & $ 0.65 $ & $ 0.61 $ & $ 0.55 $ & $ 0.66 $ & $ 0.69 $ & $ 0.62 $
& &  
$ 1.50 $ & $ 1.08 $ & $ 1.07 $ & $ 1.30 $ & $ 2.06 $ & $ 1.40 $
\\ 
K\&U 2      &  
$ 0.70 $ & $ 1.02 $ & $ 1.03 $ & $ 0.85 $ & $ 1.07 $ & $ 1.15 $ & $ 0.97 $
& &  
$ 2.93 $ & $ 1.30 $ & $ 1.12 $ & $ 1.60 $ & $ 3.28 $ & $ 2.05 $
\\       
NP      &  
$ 0.95 $ & $ 1.21 $ & $ 1.08 $ & $ 0.92 $ & $ 1.17 $ & $ 1.29 $ & $ 1.10 $
& &  
$ 2.67 $ & $ 1.39 $ & $ 1.31 $ & $ 1.60 $ & $ 3.61 $ & $ 2.12 $
\\   
%& \\ 
\multicolumn{15}{c}{Scenario 4: Self-designed; $ n_{r} = 1000 $} \\         
2 FPCs      &  
$ 0.78 $ & $ 1.24 $ & $ 1.08 $ & $ 0.89 $ & $ 1.39 $ & $ 1.26 $ & $ 1.11 $             
& &  
$ 2.68 $ & $ 1.42 $ & $ 1.27 $ & $ 1.66 $ & $ 3.59 $ & $ 2.12 $          
\\    
Adaptive      &  
$ 0.84 $ & $ 1.27 $ & $ 1.09 $ & $ 0.83 $ & $ 1.37 $ & $ 1.32 $ & $ 1.12 $  
& &  
$ 2.51 $ & $ 1.27 $ & $ 1.18 $ & $ 1.54 $ & $ 3.33 $ & $ 1.97 $          
\\                                                 
Rich      &  
$ 1.43 $ & $ 7.43 $ & $ 1.33 $ & $ 1.24 $ & $ 11.41 $ & $ 4.59 $ & $ 4.57 $    
& &  
$ 4.12 $ & $ 4.69 $ & $ 3.56 $ & $ 5.01 $ & $ 4.92 $ & $ 4.46 $         
\\
Truth      &  
$ 0.64 $ & $ 0.73 $ & $ 0.67 $ & $ 0.60 $ & $ 0.74 $ & $ 0.73 $ & $ 0.68 $        
& &  
$ 1.51 $ & $ 1.06 $ & $ 1.08 $ & $ 1.34 $ & $ 2.09 $ & $ 1.42 $          
\\
K\&U 2      &  
$ 0.85 $ & $ 1.31 $ & $ 1.28 $ & $ 1.04 $ & $ 1.37 $ & $ 1.41 $ & $ 1.21 $     
& &  
$ 3.08 $ & $ 1.36 $ & $ 1.17 $ & $ 1.78 $ & $ 3.49 $ & $ 2.18 $   
\\       
NP      &  
$ 1.15 $ & $ 1.47 $ & $ 1.29 $ & $ 1.08 $ & $ 1.40 $ & $ 1.51 $ & $ 1.32 $    
& &  
$ 2.80 $ & $ 1.40 $ & $ 1.31 $ & $ 1.69 $ & $ 3.66 $ & $ 2.17 $         
\\   
\end{tabular}
\end{adjustbox}
\label{combined_density_quan} 
\end{table}

%%%%%%%%%%%%%%%%%%%%%%%%%%%%%%%%%%%%%%%%%%%%%%%%%%%%%%%%%%%%%%%%%%%%%%

\section{Real-data analysis}
\label{sec:realdata}

In this section, we illustrate our approach using a real-world data set
that was also used by \citet {kneip2001inference}.
The data set is available on the UK Data Service website (\texttt {beta.ukdataservice.ac.uk/datacatalogue/series/series?id=200016}). 
The Family Expenditure Survey \citep[FES;][]{FES} data contain annual cross-sectional samples on the incomes and expenditure of 
more than 7000 households; we select the years 1968 to 1988.
Following \citet {kneip2001inference}, we used the log-transformed household relative income data in our data analysis.
Household relative income is the ratio of the net income to the mean income of the surveyed households,
which is often regarded as an important economic indicator for a country.

\begin{figure}%[h!]  
\centering 
\includegraphics[height = 0.5\textwidth, width = 0.5\textwidth]{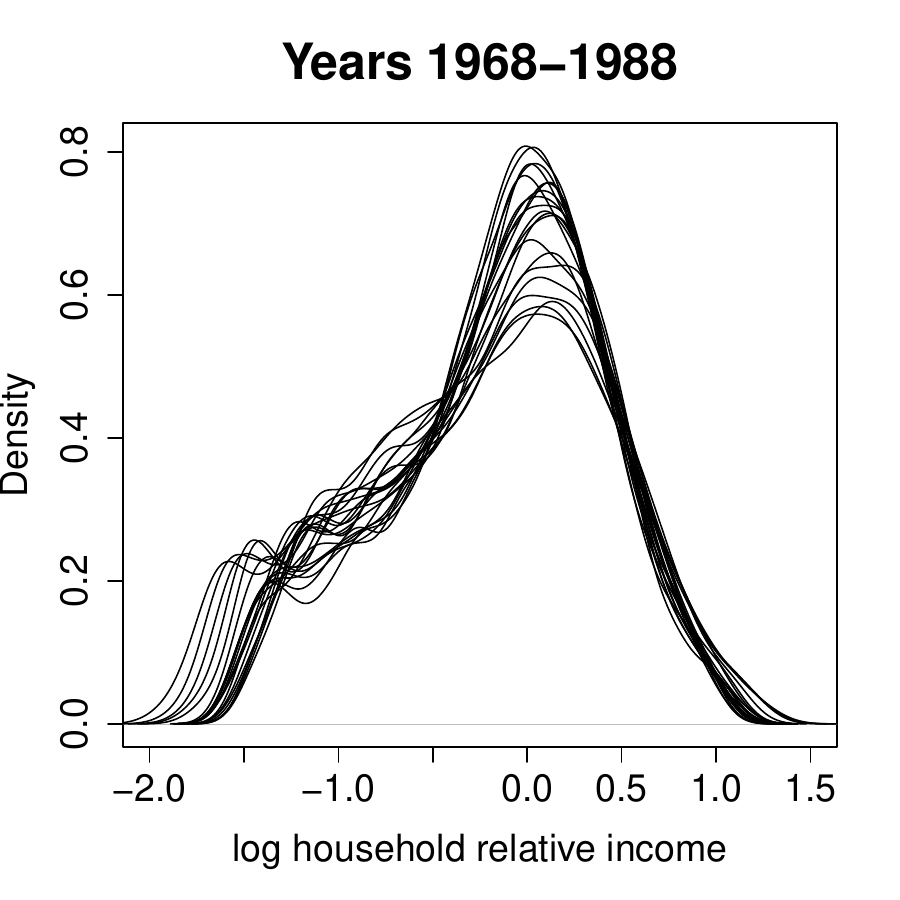} 
\caption{Kernel density estimators based on household relative income data for 1968--1988.} 
\label{KDE_plot}
\end{figure}

The raw FES data contain some outliers. 
We removed some negative values as well as $ 1\% $ of the extreme values
from the lower and upper tails.
After this simplistic cleaning, we had $ 21 $ annual samples with similar ranges.
%The cleaned data is provided in the Appendix.
See Figure~\ref {KDE_plot} for the kernel density estimates
with the Gaussian kernel and Silverman's bandwidth.
The samples are heavily right-tailed and exhibit two or more modes. 
One may therefore use finite normal or gamma mixture models for the analysis. 
The log-transformed household relative income distributions, however,
clearly have common features.
Hence, we examine the effectiveness of the inference based on the
proposed DRM with the adaptive basis function.

\begin{figure}%[h!]  
\centering 
\includegraphics[height = 0.3\textwidth, width = 0.9\textwidth]{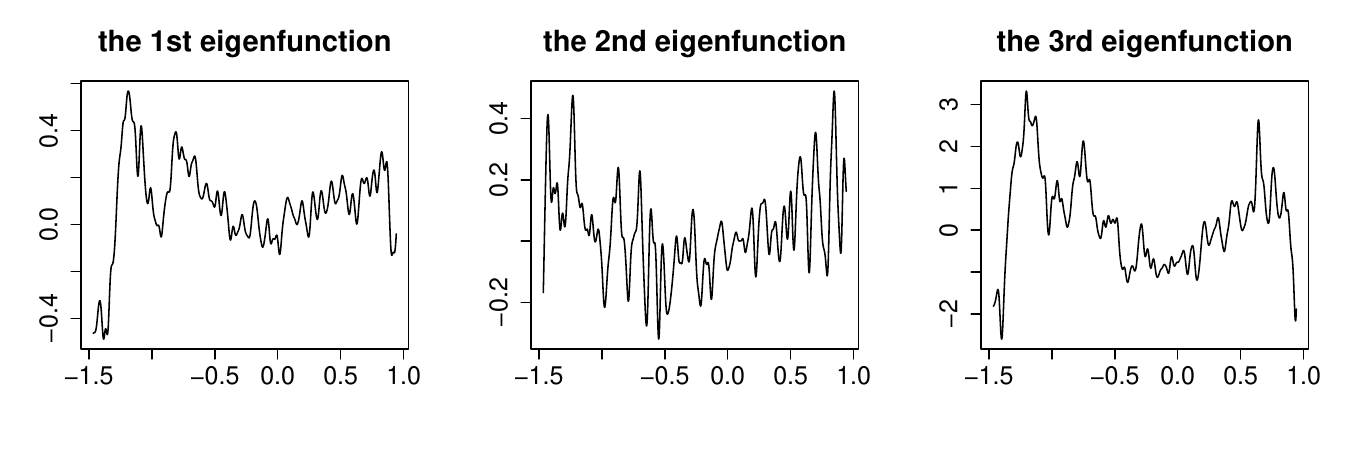} 
\caption{First three eigenfunctions obtained using data from 1968--1981.} 
\label{eigenfun_prefit_plot} 
\end{figure}

We use samples from 1968--1981 as the training data 
and samples from 1982--1988 as the test data. 
Specifically, we first obtain the adaptive basis function \eqref {adapt.basis} 
based on the training data.
We plot the first three fitted eigenfunctions in a selected range in Figure~\ref {eigenfun_prefit_plot}.
We create multiple samples by bootstrapping from the test data, and we fit the
DRM with the adaptive basis function chosen by the training data.
This procedure mimics the scenario of predicting the future aided by historical data.
We use the sample sizes $ n_{r} = 500 $ and $ 1000 $ with $ 1000 $ repetitions.
We regard the kernel density estimates and empirical quantiles
based on the full test data as the truth for the computation of the IMSE and MSE.
The remaining procedures are the same as those applied to the simulated data.
The results are given in Table~\ref {RealData_prefit_density_quan}, and
we include boxplots of these performance measures in the Appendix.

\begin{figure}%[h!]  
\centering 
\includegraphics[height = 0.25\textwidth, width = \textwidth]{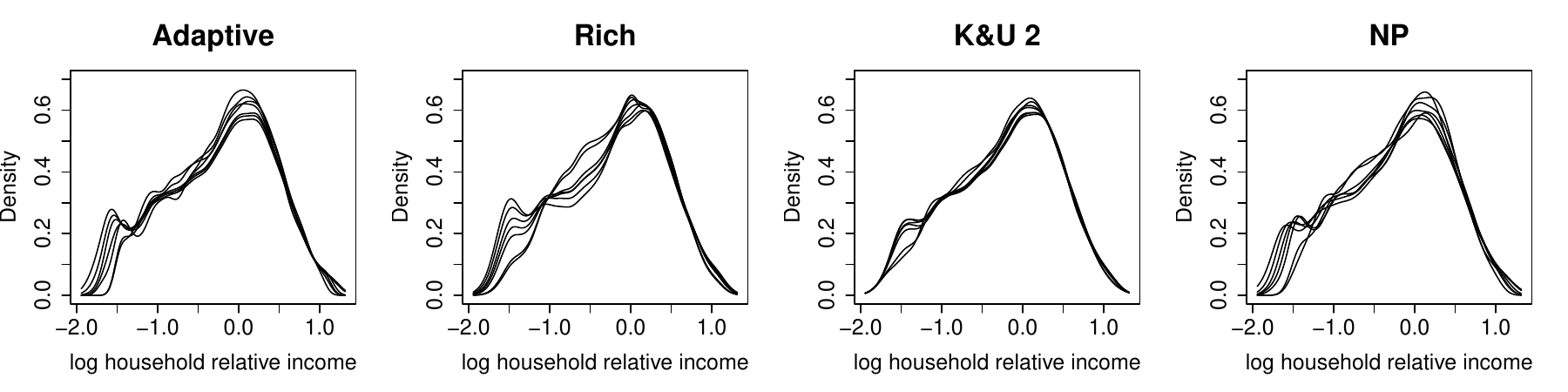} 
\caption{Estimated density functions for 1982--1988.} 
\label{density_est_plot} 
\end{figure}

The proposed DRM estimators with the fully adaptive basis functions
(``Adaptive'' in the table) outperform all the other estimators considered in this study.
On average, our estimators are $ 54\% $ and $ 38\% $ more efficient than 
the nonparametric estimators.
The estimators derived from \citet {kneip2001inference} lose efficiency
when the sample sizes are $ 1000 $. 
We find that the adaptive approach usually selects $d=2$ eigenfunctions. 
This observation supports the claim that latent structure often exists in real-world data.
In Figure~\ref {density_est_plot}, we plot the density estimators for 1982--1988 based
on some of the methods,
obtained by fitting the trained adaptive DRM and the trained model of \citet {kneip2001inference} 
to the full test data for this period.
The DRM estimates with the fully adaptive basis function
successfully characterize the similarities and differences between the distributions.
The estimates from \citet {kneip2001inference} exaggerate the similarities and are
insensitive to the differences.

\begin{table}%[h!] 
\caption{Simulated IMSEs of density estimators and average MSEs of quantile estimators scaled by respective sample sizes}
\begin{adjustbox}{max width=\textwidth}
\begin{tabular}{lrrrrrrrrrrrrrrrr}
%\hline
\multicolumn{1}{l}{Method}      
& \multicolumn{8}{c}{IMSE of density estimators}      
&
& \multicolumn{6}{c}{Average MSE of quantile estimators}       \\ 
& $ 1982 $ & $ 1983 $ & $ 1984 $ & $ 1985 $ & $ 1986 $ & $ 1987 $ & $ 1988 $ & avg.     
&
& $ 10\% $ & $ 30\% $ & $ 50\% $ & $ 70\% $ & $ 90\% $ & avg. \\ 
%\cline{2-8} \cline{10-15} 
\multicolumn{16}{c}{$ n_{r} = 500 $} \\ 
%\hline 
1 FPC      &  
$ 1.00 $ & $ 1.13 $ & $ 0.75 $ & $ 0.66 $ & $ 0.61 $ & $ 0.62 $ & $ 0.54 $ & $ 0.76 $
& &  
$ 1.21 $ & $ 0.45 $ & $ 0.26 $ & $ 0.12 $ & $ 0.20 $ & $ 0.45 $
\\         
2 FPCs      &  
$ 1.05 $ & $ 0.96 $ & $ 0.68 $ & $ 0.61 $ & $ 0.65 $ & $ 0.63 $ & $ 0.57 $ & $ 0.74 $
& &  
$ 1.03 $ & $ 0.50 $ & $ 0.30 $ & $ 0.15 $ & $ 0.27 $ & $ 0.45 $
\\      
Adaptive      &  
$ 1.05 $ & $ 0.96 $ & $ 0.68 $ & $ 0.61 $ & $ 0.65 $ & $ 0.63 $ & $ 0.57 $ & $ 0.74 $
& &  
$ 1.03 $ & $ 0.50 $ & $ 0.30 $ & $ 0.15 $ & $ 0.27 $ & $ 0.45 $
\\                             
Rich      &  
$ 2.56 $ & $ 1.60 $ & $ 1.65 $ & $ 1.45 $ & $ 1.28 $ & $ 1.39 $ & $ 2.07 $ & $ 1.71 $
& &  
$ 1.44 $ & $ 1.38 $ & $ 0.82 $ & $ 0.51 $ & $ 0.52 $ & $ 0.93 $
\\
K\&U 1      &  
$ 3.72 $ & $ 3.26 $ & $ 1.30 $ & $ 0.89 $ & $ 0.85 $ & $ 1.34 $ & $ 1.74 $ & $ 1.87 $
& &  
$ 4.71 $ & $ 1.92 $ & $ 0.56 $ & $ 0.13 $ & $ 0.90 $ & $ 1.64 $
\\ 
K\&U 2      &  
$ 2.30 $ & $ 1.80 $ & $ 1.51 $ & $ 1.08 $ & $ 0.87 $ & $ 1.10 $ & $ 1.61 $ & $ 1.47 $
& &  
$ 2.20 $ & $ 0.88 $ & $ 0.41 $ & $ 0.20 $ & $ 0.73 $ & $ 0.88 $
\\            
NP      &  
$ 1.62 $ & $ 1.57 $ & $ 1.72 $ & $ 1.66 $ & $ 1.57 $ & $ 1.49 $ & $ 1.54 $ & $ 1.60 $
& &  
$ 1.73 $ & $ 1.46 $ & $ 0.83 $ & $ 0.56 $ & $ 0.69 $ & $ 1.05 $
\\   
\multicolumn{16}{c}{$ n_{r} = 1000 $} \\ 
%\hline 
1 FPC      &  
$ 1.57 $ & $ 1.96 $ & $ 1.09 $ & $ 0.98 $ & $ 1.01 $ & $ 1.05 $ & $ 0.83 $ & $ 1.21 $
& &  
$ 1.86 $ & $ 0.62 $ & $ 0.37 $ & $ 0.16 $ & $ 0.31 $ & $ 0.66 $
\\         
2 FPCs      &  
$ 1.59 $ & $ 1.57 $ & $ 0.98 $ & $ 0.88 $ & $ 1.06 $ & $ 1.06 $ & $ 0.87 $ & $ 1.14 $
& &  
$ 1.43 $ & $ 0.68 $ & $ 0.44 $ & $ 0.22 $ & $ 0.40 $ & $ 0.63 $
\\    
Adaptive      &  
$ 1.59 $ & $ 1.57 $ & $ 0.98 $ & $ 0.88 $ & $ 1.06 $ & $ 1.06 $ & $ 0.88 $ & $ 1.15 $
& &  
$ 1.43 $ & $ 0.69 $ & $ 0.44 $ & $ 0.22 $ & $ 0.40 $ & $ 0.64 $
\\                             
Rich      &  
$ 4.02 $ & $ 2.28 $ & $ 2.24 $ & $ 1.82 $ & $ 1.58 $ & $ 1.97 $ & $ 3.57 $ & $ 2.50 $
& &  
$ 1.88 $ & $ 1.39 $ & $ 0.91 $ & $ 0.56 $ & $ 0.54 $ & $ 1.06 $ 
\\
K\&U 1      &  
$ 6.60 $  &  $ 5.80 $  &  $ 1.93 $  &  $ 1.30 $  &  $ 1.49 $  &  $ 2.64 $  &  $ 3.56 $ & $ 3.33 $
& &  
$ 9.03 $ & $ 3.70 $ & $ 0.96 $ & $ 0.20 $ & $ 1.28 $ & $ 3.03 $
\\ 
K\&U 2      &  
$ 3.55 $  &  $ 2.83 $  &  $ 2.14 $  &  $ 1.39 $  &  $ 1.18 $  &  $ 1.78 $  &  $ 2.91 $ & $ 2.26 $ 
& &  
$ 3.82 $ & $ 1.19 $ & $ 0.62 $ & $ 0.31 $ & $ 0.89 $ & $ 1.36 $
\\        
NP      &  
$ 1.86 $ & $ 1.87 $ & $ 1.99 $ & $ 1.91 $ & $ 1.85 $ & $ 1.75 $ & $ 1.70 $ & $ 1.85 $
& &  
$ 1.78 $ & $ 1.41 $ & $ 0.84 $ & $ 0.57 $ & $ 0.67 $ & $ 1.05 $ 
\\   
%\hline 
\end{tabular}
\end{adjustbox}
\label{RealData_prefit_density_quan} 
\end{table}

%%%%%%%%%%%%%%%%%%%%%%%%%%%%%%%%%%%%%%%%%%%%%%%%%%%%%%%%%%%%%%%%%%%%%%

\section*{Acknowledgement}
This research work is supported in part by Natural Science and Engineering Research Council of Canada and FPInnovations.
The authors would also like to thank Professor Nancy Heckman and Professor James Zidek for helpful discussions and support.

\appendix
\section*{Appendix} 
\label{app} 

This Appendix includes proofs of the technical results and provides additional simulation results and figures. 
The additional experiments answer another important question: 
does the DRM perform well when the samples are from populations whose distributions are not connected
through a few basis functions? The answer is positive, based on simulations with
data sets generated from the Weibull and finite normal mixture distributions.
We also include all the boxplots here to allow a more complete comparison of the various approaches. 
We will generally use the notation of the main paper.

\subsection*{Proof of Theorem~\ref {thm.MV}}
\noindent
We show that the eigenvectors of $\bM$ and eigenfunctions of
the operator $\Vcal$ are connected as stated in this theorem.
Let
\[
\bQ (t) = (Q_0^c(t), \ldots, Q_m^c(t))^\top
\]
so that we may write $ \Sigma (s, t) = \bQ^\top (s) \bQ (t) $ and 
$ \bM = \int \bQ (t) \bQ^{\top} (t) \mathrm {d} \bar G (t)$.
Define a matrix of eigenvectors and a diagonal matrix of eigenvalues of $ \bM $:
\begin{gather*}
\bP = [\bp_0, \ldots, \bp_{d-1}];  \\
\bLambda = \mbox{diag} (\lambda_0, \ldots, \lambda_{d-1} ).
\end{gather*}
By the definition of eigenvalue and eigenvector, we have $\bM \bP = \bP \bLambda$. 
Next, we form a vector of possible eigenfunctions of $\Vcal$:
\[
\bPsi (t) = (\psi_0 (t), \ldots, \psi_{d-1} (t))^\top 
= \bLambda^{-1/2} \bP^{\top} \bQ (t).
\]
With this preparation, we have
\ba
\Vcal \bPsi (t)
&=& 
\int \bPsi (s) \Sigma (s, t) \mathrm {d} \bar G (s) \\
&=& 
\int \bLambda^{-1/2} \bP^{\top} \bQ (s) \bQ^\top (s) \bQ (t) \mathrm {d} \bar G (s) \\
&=&
\bLambda^{-1/2} \bP^{\top} \Big \{ \int \bQ (s) \bQ^\top (s) \mathrm {d} \bar G (s) \Big \} \bQ (t) \\ 
&=&
\bLambda^{-1/2} \{ \bP^{\top} \bM \} \bQ (t) \\ 
&=&
\bLambda^{-1/2} \bLambda \bP^{\top} \bQ (t) \\ 
&=&
\bLambda \bPsi (t).
\ea
That is, $ \Vcal \bPsi (t) = \bLambda \bPsi (t) $, 
which implies that $ \bPsi (t)$ is indeed a vector of eigenfunctions with the eigenvalues 
$ \lambda_0, \ldots, \lambda_{d-1} $.
This proves the first property stated in the theorem.

We next notice that
\[
\int \bPsi (s) \bPsi^\top (s) \mathrm {d} \bar G (s) 
= \bLambda^{-1/2} \bP^\top \bM \bP \bLambda^{-1/2} = \bI,
 \]
an identity matrix. This result confirms that the elements of $ \bPsi (t) $ are orthonormal.
This proves the second property stated in the theorem.

Finally, we note that the linear space spanned by $ \{ \bPsi_j (t), j=0, \ldots, d-1 \} $
is a subspace of the space spanned by the elements of $ \bq (t) $, 
and the ranks of both spaces are $ d $. 
Therefore, these two spaces must be identical.
This proves the third property of the theorem
and completes the proof.

\vspace{1ex}

\subsection*{Proof of Theorem~\ref {consist_M}}
This theorem states that $ \hat \bM $ is consistent for $ \bM $. 
The consistency of the eigenvalues and eigenvectors are consequences.
Recall that for $0 \leq i, j \leq m$, 
\ba
\bM(i, j) 
& = & 
\bar E \{ \int Q^c_i(X) Q^c_j(X)\} 
= \int Q^c_i (x) Q^c_j (x) \mathrm {d} \bar G (x); \\
\hat {\bM}({i, j} )
& = &
\int \hat {Q}^c_i (x) \hat {Q}^c_j (x) \mathrm {d} \bar F_{n} (x). 
\ea
We prove this theorem by showing that for any $(i, j)$ in this range, 
\[
\hat {\bM} (i, j) \to \bM (i, j),
\]
in probability,
as the total sample size $N \to \infty$ under the DRM
and  under the model conditions specified in the theorem.

By the moment condition $ \bar E | \log g_{i} (X) |^{3} < \infty $, it can be seen
that
$Q^c_i(X)$ has a finite second moment when $X$ has distribution $\bar G$.
By the classical law of large numbers,
\[
 \int Q^c_i (x) Q^c_j (x) \mathrm {d}  \bar F_{n} (x) 
 \to \int Q^c_i (x) Q^c_j (x) \mathrm {d} \bar G (x)
\]
almost surely. Hence, we need only show that
\begin{align}
\label{eqnA1}
\int \{  \hat {Q}^c_i (x)  \hat {Q}^c_j (x)  - Q^c_i(x) Q^c_j(x)\} \mathrm {d} \bar F_{n} (x)
\to 0
\end{align}
in probability.
Let us decompose this difference into
\begin{align}
\label{eqnA2}
& \hspace*{-2em} \hat Q^c_i (x) \hat Q^c_j (x) - Q^c_i (x) Q^c_j (x) \\
&=
 \{ \hat Q^c_i (x)  - Q^c_i (x) \} \{ \hat Q^c_j (x) - Q^c_j (x)\}
 \nonumber \\
 & \hspace*{1em} 
 +
 Q^c_i (x) \{ \hat Q^c_j (x) - Q^c_j (x) \}
 +
Q^c_j (x) \{ \hat Q^c_i (x) - Q^c_i (x) \}.
\nonumber
\end{align}
By Cauchy's inequality, we have a bound for the second term on the
right-hand side of \eqref{eqnA2}:
\[
\Big \{ \int {Q}^c_i (x)  \{ \hat {Q}^c_j (x)  - Q^c_j(x) \}  \mathrm {d} \bar F_{n} (x) \Big \}^2
\leq
 \int \{ {Q}^c_i (x) \}^2  \mathrm {d} \bar F_{n} (x) 
 \int \{ \hat {Q}^c_j (x)  - Q^c_j(x) \}^2 \mathrm {d} \bar F_{n} (x).
\]
Under the moment condition, the first multiplication factor in the above expression has a finite limit by the law of
large numbers. Thus, if 
\begin{align}
\label{eqnA3}
 \int \{ \hat {Q}^c_j (x)  - Q^c_j(x) \}^2 \mathrm {d} \bar F_{n} (x)   \to 0
\end{align}
 in probability, then
 \[
 \int {Q}^c_i (x)  \{ \hat {Q}^c_j (x)  - Q^c_j(x) \} \mathrm {d} \bar F_{n} (x) \to 0.
\]
We can use similar bounds from Cauchy's inequality for the other terms in \eqref{eqnA2}
to arrive at the same conclusion. These lead to \eqref{eqnA1}.
Hence, our task is reduced to proving \eqref{eqnA3}.
 
We first show that for each $i$,
 \begin{align}
 \label{eqnA4}
 \int \{ \log \hat g_i(x) - \log g_i(x) \}^2 \mathrm {d} \bar F_{n} (x)   \to 0.
\end{align}
Let $\Xcal_{\delta, i} = \{x:  g_i (x) \leq \delta \}$ with $\delta = N^{-1/3}$ to simplify the notation.
We can see that
 \begin{align*}
\int_{\Xcal_{\delta, i} } | \log g_i (x) |^2 \mathrm {d} \bar F_{n} (x)
&\leq
\int_{\Xcal_{\delta, i} } \frac{3 | \log g_i (x) |^3}{\log N} \mathrm {d} \bar F_{n} (x)\\
&\leq
3 \log ^{-1} N \int | \log g_i (x) |^3 \mathrm {d} \bar F_{n} (x)
\to 0,
\end{align*}
since the second factor is $ O_p (1) $ under the moment condition 
$ \bar E | \log g_{i} (X) |^{3} < \infty $.

Since  $\hat g_i(x)$ is uniformly consistent for $g_i(x)$,
without loss of generality, we assume $\hat g_i(x) < 1$ to simplify the presentation.
Recall that $\hat g_i(x) \geq C_i (\log N /N)^{2/5}$ by design,
so we get
\[
| \log \hat g_i (x) |^2 \leq \log^2 N,
\]
when $N$ is large enough. We have removed some harmless constants to simplify the expression.
Note that by a well-known result about the empirical distribution, we have
$\sup_x | \bar F_n(x) - \bar{G} (x)| = O( (\log N/N)^{1/2})$.
Hence,
\[
\int_{\Xcal_{\delta, i} } | \log \hat g_i (x) |^2 \mathrm {d} \bar F_{n} (x) 
\leq
 (\log^2 N ) \bar{F}_n\{ \Xcal_{\delta, i} \}
 = O_p( (\log N)^{-1} )
= o_p(1).
\]

Combining two order assessments of the integration
of $\log g_i(x)$ and $\log \hat g_i(x)$ on ${\Xcal_{\delta, i} }$,
we arrive at a partial conclusion of  \eqref{eqnA4}:
\[
 \int_{\Xcal_{\delta, i}} \{ \log \hat g_i (x) -  \log g_i (x) \}^2 \mathrm {d} \bar F_{n} (x) \to 0
\]
in probability. We now use Lemma~\ref {lemma_kernel}.
Let ${\Xcal}^c_{\delta,i}$ be the complement of $\Xcal_{\delta,i}$.
With $ \delta = N^{-1/3} $ and $ \sup_x | \hat g_i (x) - g_i (x)| = O( (\log N/N)^{2/5} )$,
we have 
\[
\inf_{x \in {\Xcal}^c_{\delta,i}} \hat g_i (x) = N^{-1/3} (1+ o(1)).
\]
For simplicity, we proceed as if $ \inf_{x \in {\Xcal}^c_{\delta,i}} \hat g_i (x) = N^{-1/3} $ so that
\[
\sup_{x \in {\Xcal}^c_{\delta,i}} | \log \hat g_i (x) - \log g_i (x) |^2
\leq 
N^{2/3} \sup_{x \in {\Xcal}^c_{\delta,i}} | \hat g_i (x) - g_i (x) |^2
= o(1).
\]
This implies
\[
 \int_{\Xcal^c_{\delta, i}} | \log \hat g_i (x) - \log g_i (x) |^2 \mathrm {d} \bar F_n(x)
 = o(1).
\]
This completes the proof of \eqref{eqnA4}.

Note that $\hat Q_i^c (x) - Q_i^c (x)$ is composed of
various linear combinations of $ \log \hat g_i (x) - \log g_i (x) $.
We can easily see that \eqref{eqnA4} implies \eqref{eqnA3}.
We omit these tedious details here because they are conceptually simple.

\subsection*{Proof of Theorem~\ref {consist_eigen}} 

This theorem states that the eigenvalues and eigenvectors of $ \hat \bM $ are consistent for those of $ \bM $.
Because the eigenvalues and eigenvectors are continuous
functions of $ \bM $, and $ \hat \bM $ is consistent for $ \bM $
as proved in the previous section,
the conclusions are obvious when all the eigenvalues are different. 
When some nonzero eigenvalues are equal, their eigenvectors
form a linear space that is continuous with respect to $ \bM $.
The consistency of subspaces suffices.

We have shown in the previous section, as a side fact, that $ \hat Q^c_r(x) $ is
consistent for $ Q^c_r (x) $. Therefore, the consistency of
$ \hat \psi_j (x) $ is obvious by inspecting its definition.
Interested readers are referred to a useful lemma in \citet[Lemma A.1] {kneip2001inference}
for details regarding the relationship between the eigensystem of a matrix $ A $ 
and the eigensystem of a perturbed matrix $ A + B $.

\subsection*{Additional figures}

We include here some additional figures for the simulation studies 
and real-data analysis in the main text.

\begin{figure}%[h!] 
\centering 
\caption{Density functions for the six self-designed distributions.}
\label {BS1_true_density_plot}
\includegraphics[height = 8cm, width = 12cm]{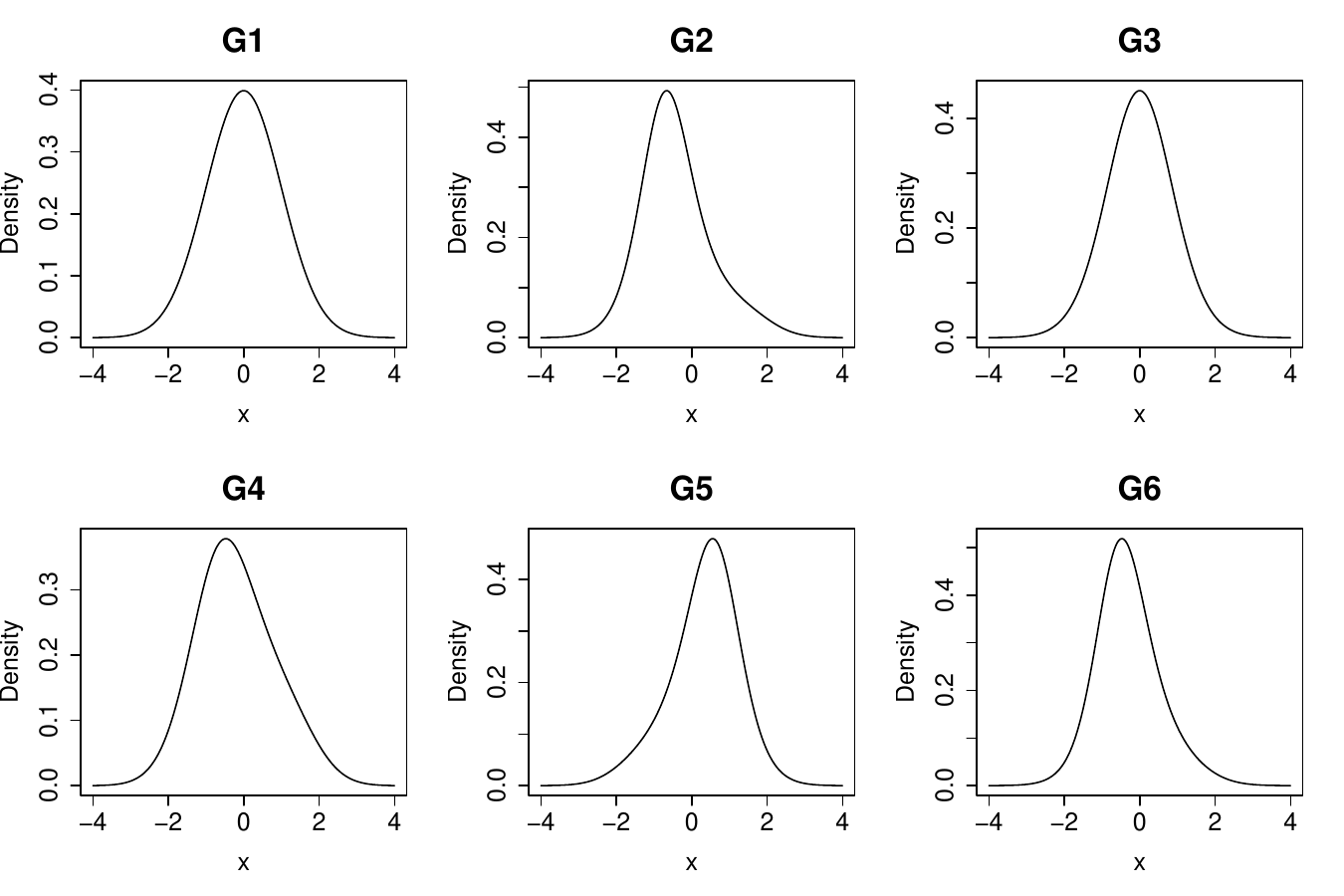} 
\end{figure}

Figure~\ref {BS1_true_density_plot} presents the six density functions of {\bf Scenario 4},
which satisfy the DRM with tailor-designed log density ratios.
These distributions have similar shapes, but the 
commonly used basis functions under DRM do not work. This example
highlights the need for adaptive basis functions.

\begin{figure}%[h!] 
\centering 
\caption{Boxplots of simulated ISEs of density estimators 
and simulated SEs of quantile estimators. 
Data generated from normal distributions with equal variances (Scenario 1).}
\label {Normal1_boxplots_n1000}
\includegraphics[height = 20cm, width = \textwidth]{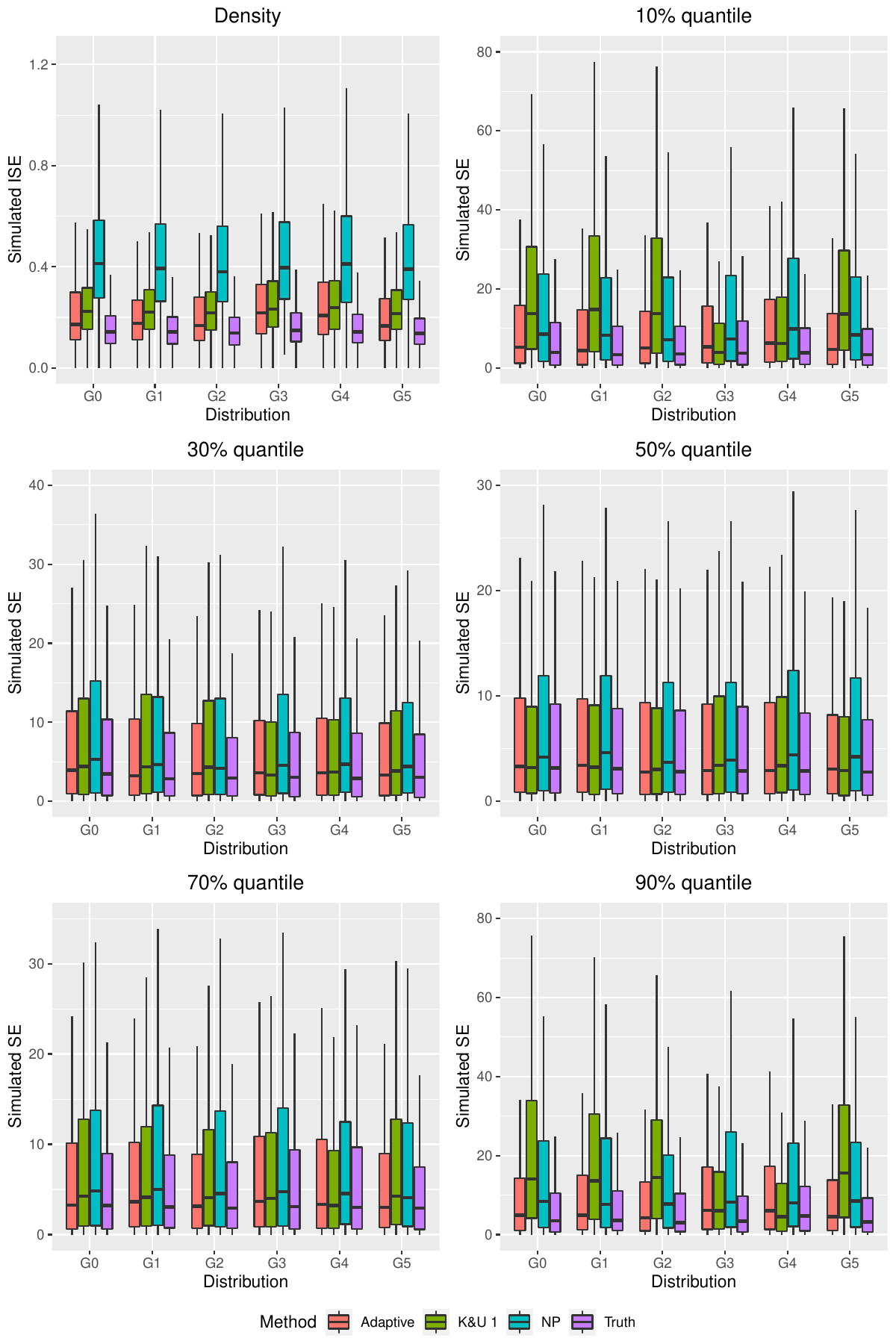} 
\end{figure}

\begin{figure}%[h!] 
\centering 
\caption{Boxplots of simulated ISEs of density estimators 
and simulated SEs of quantile estimators. 
Data generated from normal distributions with unequal variances (Scenario 2).}
\label {Normal2_boxplots_n1000}
\includegraphics[height = 20cm, width = \textwidth]{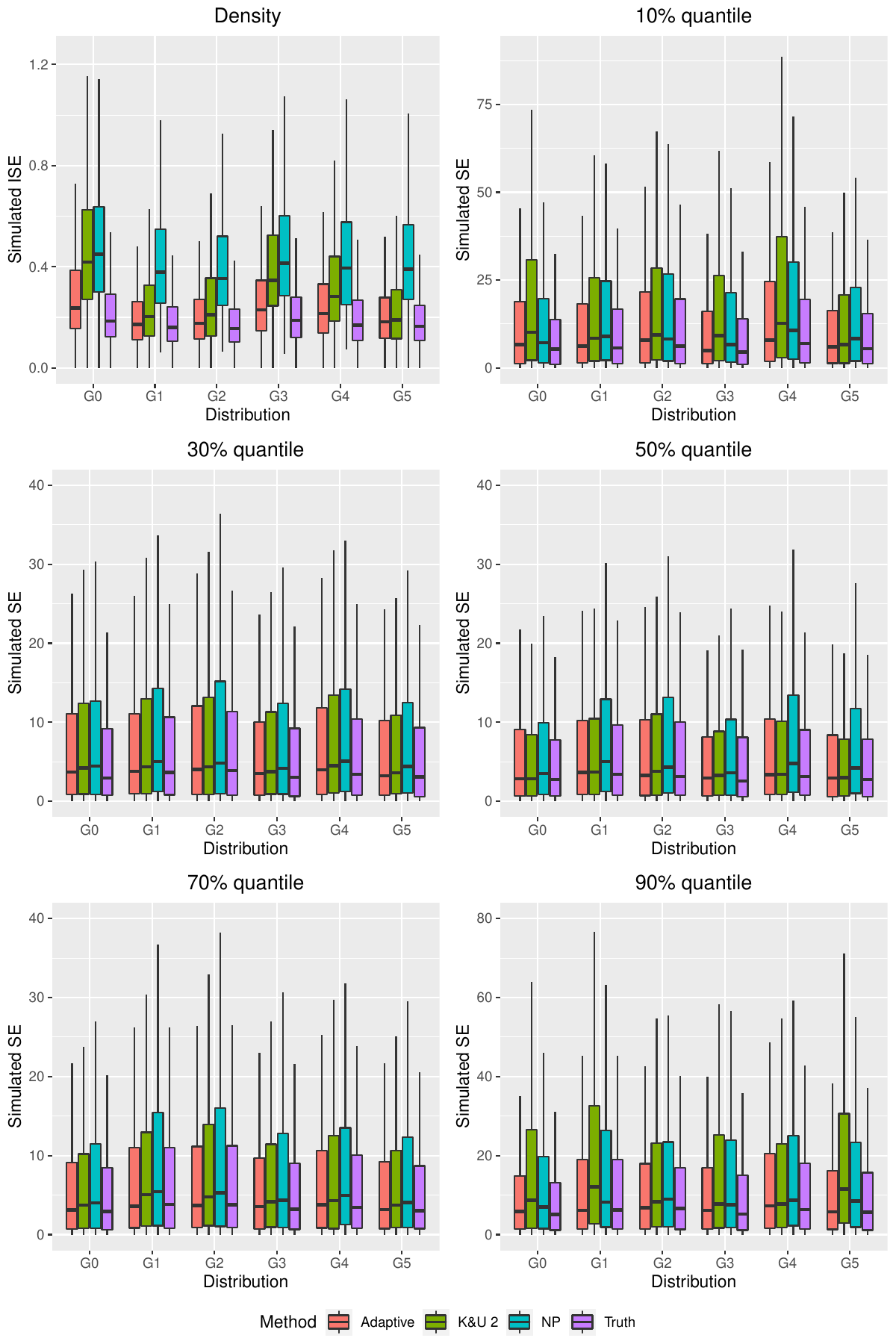} 
\end{figure}

\begin{figure}%[h!] 
\centering 
\caption{Boxplots of simulated ISEs of density estimators 
and simulated SEs of quantile estimators. 
Data generated from gamma distributions (Scenario 3).}
\label {Gamma1_boxplots_n1000}
\includegraphics[height = 20cm, width = \textwidth]{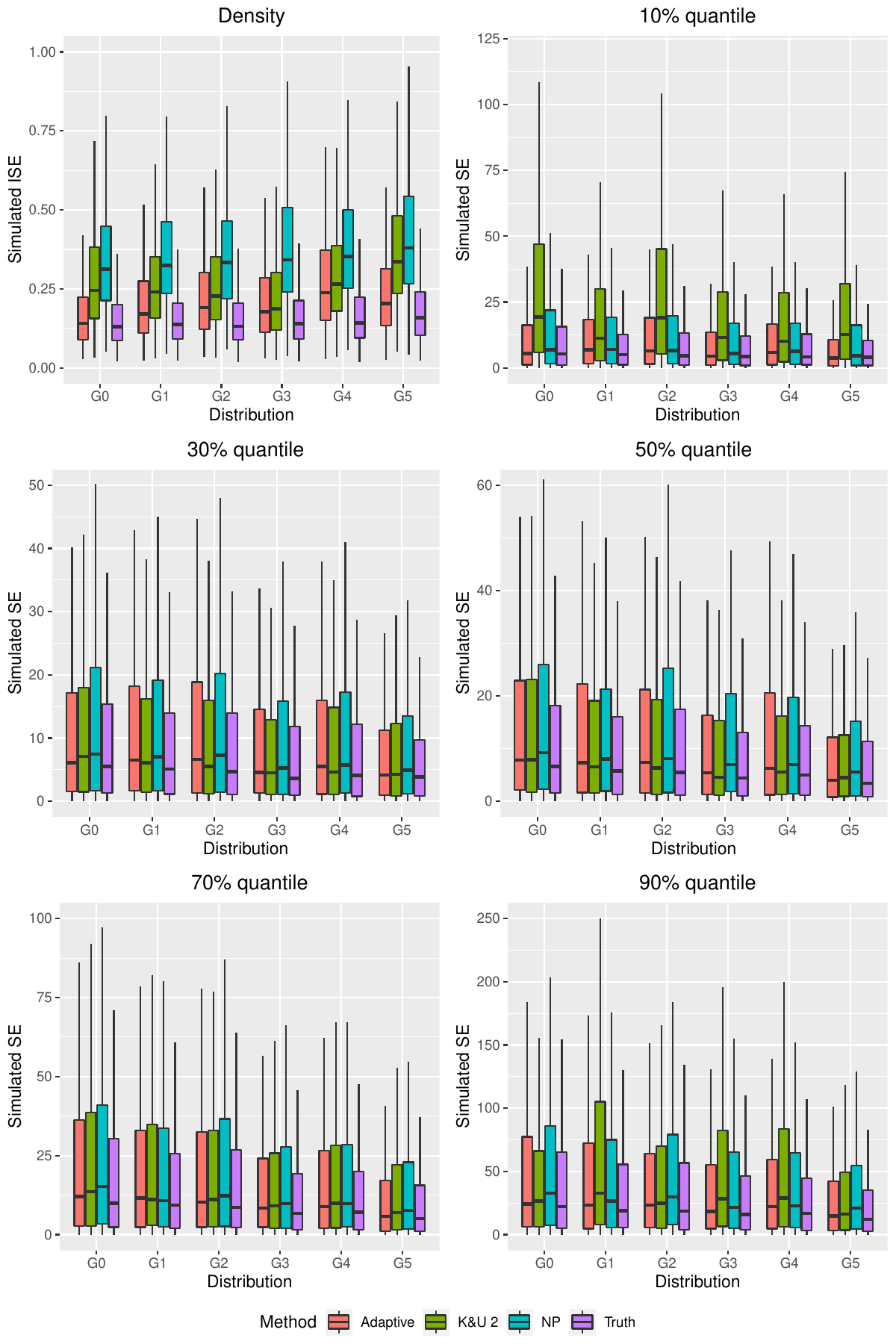} 
\end{figure}

\begin{figure}%[h!] 
\centering 
\caption{Boxplots of simulated ISEs of density estimators 
and simulated SEs of quantile estimators. 
Data generated from self-designed distributions (Scenario 4).}
\label {BS1_boxplots_n1000}
\includegraphics[height = 20cm, width = \textwidth]{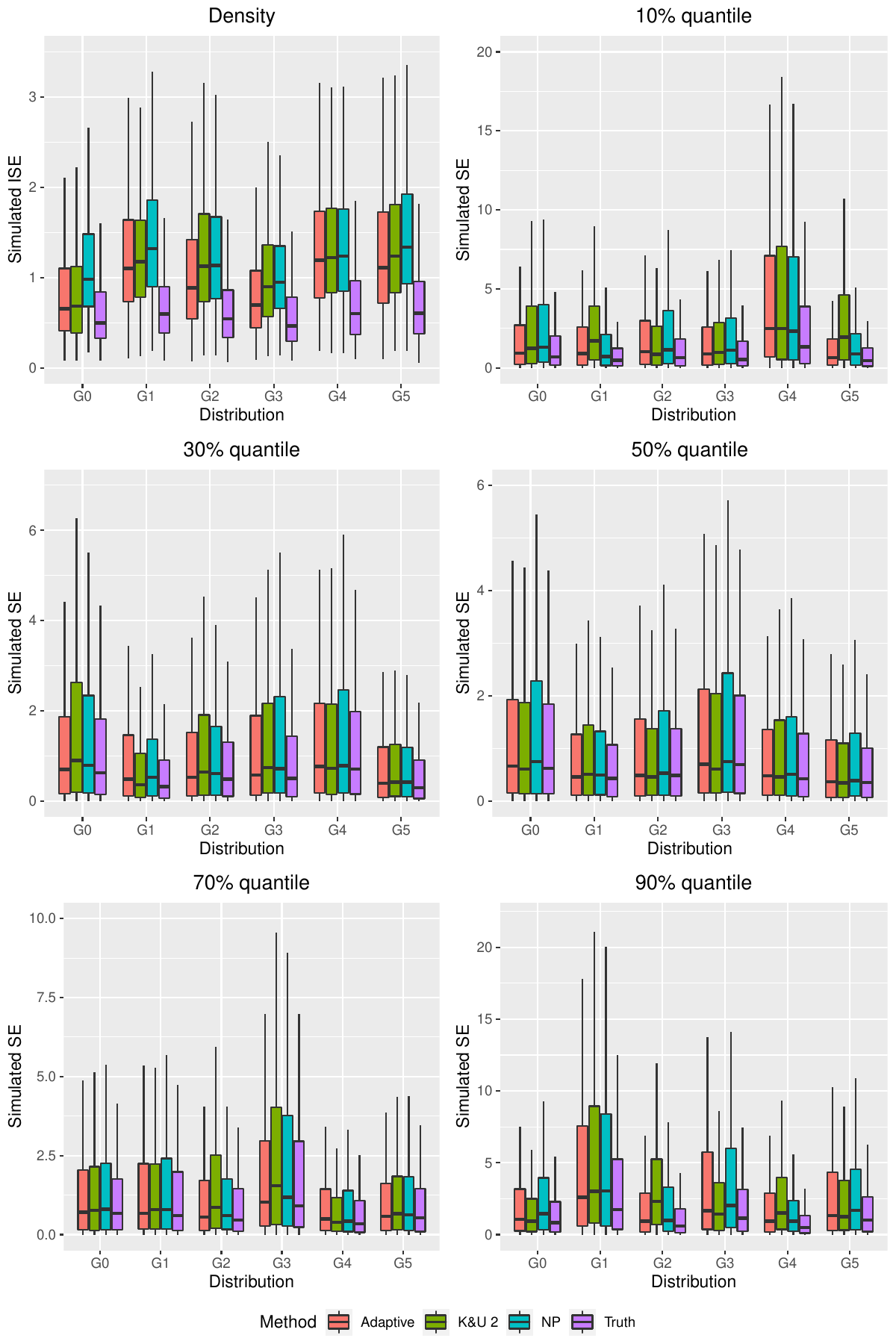} 
\end{figure}

\begin{figure}%[h!] 
\centering 
\caption{Boxplots of simulated ISEs of density estimators 
and simulated SEs of quantile estimators. 
Data from the real-data example.}
\label {RealData_prefit_boxplots_n1000}
\includegraphics[height = 20cm, width = \textwidth]{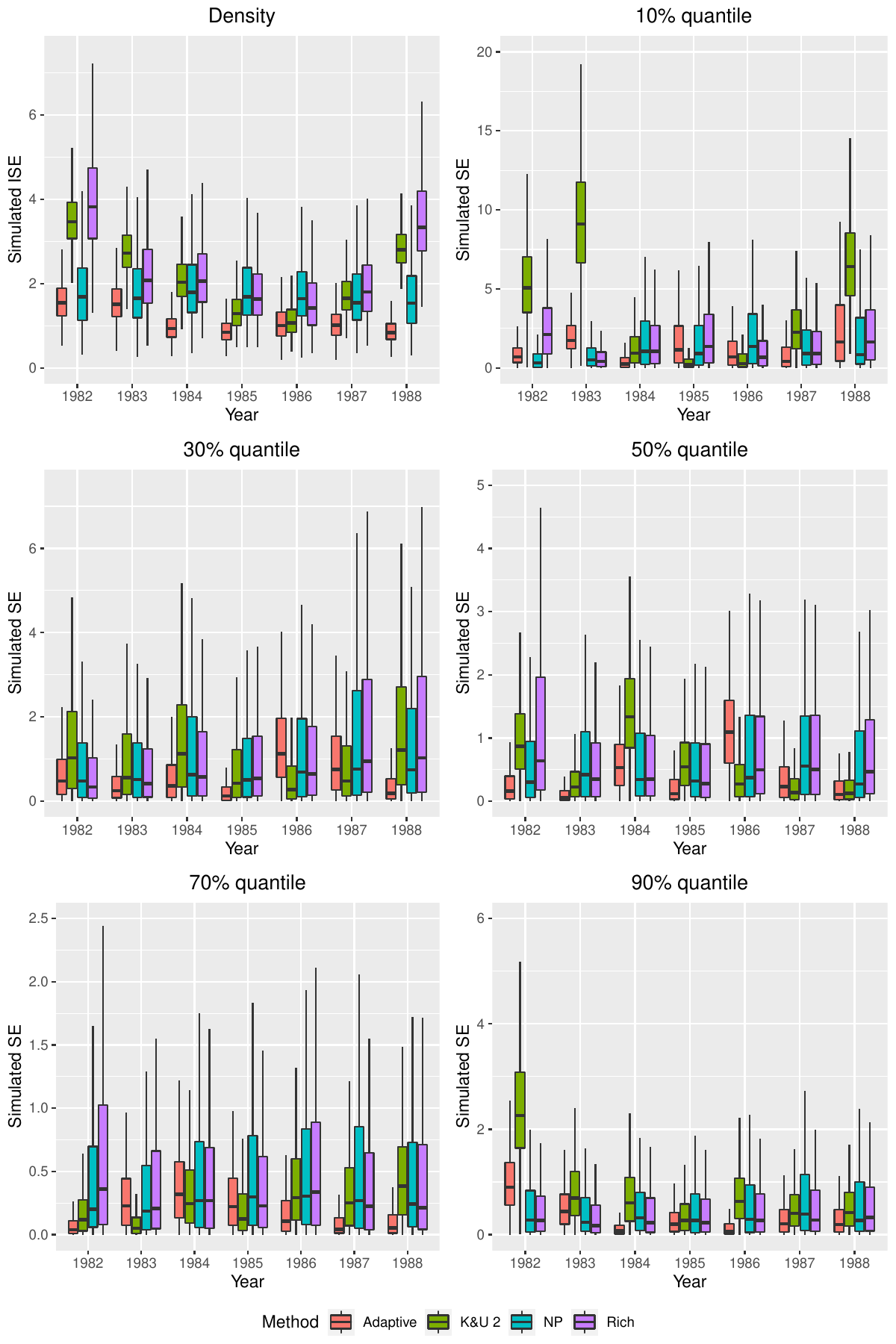} 
\end{figure}

Figures~\ref {Normal1_boxplots_n1000}--\ref {BS1_boxplots_n1000} 
present boxplots of two performance measures:
the integrated squared errors (ISE) of the density estimators and the 
simulated squared errors (SE) of the quantile estimators.
That is, these values are not averaged over $ N_{\mathrm {rep}}=1000 $ repetitions.
The boxplots shed light on the sample distributions of these
estimators, rather than merely their biases and variances.
We include the estimators for the samples from
the four scenarios presented in the main text.
Figure~\ref {RealData_prefit_boxplots_n1000} shows the boxplots 
of these two measures in the real-data example.
The sample size is $ 1000 $, and we have removed extreme values for clearer illustrations.

Our method is not always the best, but its performance is often close to that of
the DRM with the ``true'' basis function.
It outperforms the nonparametric methods in almost every case, and it usually
outperforms the method of \citet{kneip2001inference}. 

\subsection*{Simulations based on multiple samples from distributions that do not satisfy DRM} 

While data from multiple populations of a similar nature should share some
latent structure, there is no guarantee that there is always
a helpful DRM with a meaningful basis function $ \bq (x) $.
Some studies suggest that a DRM coupled with a rich basis function such as 
$ \bq (x) = (|x|^{1/2}, x, x^2, \log (1+|x|))^{\top} $ will work well. 
It is likely that a rich basis function will often but not always permit a good fit of the DRM to the data.
In this section, we present simulation results based on
data generated from Weibull distributions and finite mixture distributions.
There is no apparent DRM with a sensible choice of basis function that
includes these distributions.
One may saturate the density ratio space to ensure validity, 
but such a DRM does not truly help in terms of connecting these populations.

We fit the data with the DRM using the proposed adaptive basis function with varying numbers of eigenfunctions.
We also implement the method of \citet {kneip2001inference} and fit a DRM coupled with a rich basis function.
We wish to discover which approach will lead to consistent efficiency gains in the density 
and quantile estimation. 

We first consider $ m+1=6 $ samples generated from 
six Weibull distributions. 
The probability density function of a Weibull random variable is
\[ 
f (x; \lambda, k) = \frac {k} {\lambda} (x/\lambda)^{k-1} \exp \{ -(x/\lambda)^{k} \}, 
\hspace {5mm}
x \geq 0, 
\] 
where $ k, \lambda $ are positive values known as the shape and scale parameters. 
We set the shape parameters to $ (4.5, 5, 6, 6.5, 7, 7.5) $ and 
the scale parameters to $ (10, 9, 11, 11.5, 12.5, 12) $.

We next consider $ m+1=6 $ samples from six
two-component normal mixture distributions.
These distributions are as follows: 
\begin{gather*} 
G_{j} = \lambda_{j} N (\mu_{1, j}, \sigma_{1, j}^{2}) 
+ (1 - \lambda_{j}) N (\mu_{2, j}, \sigma_{2, j}^{2}), 
\hspace {5mm} 
j = 0, 1, \ldots, 5. 
\end{gather*} 
We choose the six mixture distributions as follows.
We set the mixing proportions to 
$ (\lambda_{0}, \ldots, \lambda_{5}) = (0.5, 0.5, 0.3, 0.5, 0.4, 0.2) $;
the component means to 
$ (\mu_{1, 0}, \ldots, \mu_{1, 5}) = (15, 15, 15.5, 16, 14.5, 15) $ 
and 
$ (\mu_{2, 0}, \ldots, \mu_{2, 5}) = (18, 18, 17, 19, 17, 16) $;
and the component variances to
$ (\sigma_{1, 0}^{2}, \ldots, \sigma_{1, 5}^{2}) = (3, 3, 3.5, 3, 3.5, 3) $
and
$ (\sigma_{2, 0}^{2}, \ldots, \sigma_{2, 5}^{2}) = (3, 3.5, 4, 4, 4.5, 3) $.

In both scenarios, we generate $ 1000 $ sets of independent and multiple
samples of sizes $ n_{r} = 500, 1000 $.
Table~\ref {nonDRM_combined_density_quan} gives 
the IMSEs of the density estimators 
and the average MSEs of the quantile estimators 
across the five quantile levels 10\%, 30\%, 50\%, 70\%, and 90\%. 
The abbreviations used in this table are the same as those in the main text.
Based on the simulation results, we find that the
DRM-based estimators have a worthwhile gain in efficiency
compared to purely nonparametric data analysis.
The use of a single eigenfunction should be avoided.
The fully adaptive basis functions work well in most cases, and the
rich basis function seems to have the best performance.
If one uses sufficient orthonormal functions in
the method of \citet {kneip2001inference}, the resulting
estimates are in line with the nonparametric
estimates.

In summary, for these two examples, applying the DRM
with either a fully adaptive basis function or
the rich basis function can lead to efficiency gains.
In applications, one cannot know if the rich basis will work well, 
and therefore we recommend the fully adaptive DRM approach.

\begin{table}%[h!] 
\caption{Simulated IMSEs of density estimators and average MSEs of quantile estimators scaled by respective sample sizes.}
\begin{adjustbox}{max width=\textwidth}
\begin{tabular}{lrrrrrrrrrrrrrrr}
%\hline
\multicolumn{1}{l}{Method}      
& \multicolumn{7}{c}{IMSE of density estimators}      
&
& \multicolumn{6}{c}{Average MSE of quantile estimators}       \\ 
& $ G_0 $ & $ G_1 $ & $ G_2 $ & $ G_3 $ & $ G_4 $ & $ G_5 $ & avg.      
&
& $ 10\% $ & $ 30\% $ & $ 50\% $ & $ 70\% $ & $ 90\% $ & avg. \\ 
%& \\ 
\multicolumn{15}{c}{$ n_{r} = 500 $, Weibull} \\ 
1 FPC      &  
$ 0.67 $ & $ 1.23 $ & $ 0.31 $ & $ 0.31 $ & $ 0.56 $ & $ 0.73 $ & $ 0.64 $
& &  
$ 31.58 $ & $ 11.77 $ & $ 5.91 $ & $ 7.19 $ & $ 21.89 $ & $ 15.67 $
\\         
2 FPCs      &  
$ 0.32 $ & $ 0.51 $ & $ 0.28 $ & $ 0.29 $ & $ 0.34 $ & $ 0.35 $ & $ 0.35 $
& &  
$ 12.52 $ & $ 7.14 $ & $ 5.45 $ & $ 4.87 $ & $ 6.33 $ & $ 7.26 $
\\    
Adaptive      &  
$ 0.43 $ & $ 0.70 $ & $ 0.30 $ & $ 0.29 $ & $ 0.43 $ & $ 0.45 $ & $ 0.43 $
& &  
$ 18.11 $ & $ 8.49 $ & $ 5.51 $ & $ 5.36 $ & $ 10.40 $ & $ 9.57 $
\\                             
Rich      &  
$ 0.33 $ & $ 0.35 $ & $ 0.27 $ & $ 0.28 $ & $ 0.33 $ & $ 0.34 $ & $ 0.32 $
& &  
$ 11.82 $ & $ 6.52 $ & $ 5.08 $ & $ 4.77 $ & $ 5.70 $ & $ 6.78 $
\\
K\&U 1      &  
$ 0.21 $ & $ 0.69 $ & $ 2.29 $ & $ 1.27 $ & $ 2.21 $ & $ 0.77 $ & $ 1.24 $
& &  
$ 49.69 $ & $ 13.32 $ & $ 11.72 $ & $ 23.57 $ & $ 52.68 $ & $ 30.20 $
\\ 
K\&U 2      &  
$ 0.29 $ & $ 0.41 $ & $ 0.44 $ & $ 0.30 $ & $ 0.45 $ & $ 0.66 $ & $ 0.43 $
& &  
$ 17.76 $ & $ 7.14 $ & $ 5.01 $ & $ 5.08 $ & $ 10.64 $ & $ 9.12 $
\\       
NP      &  
$ 0.42 $ & $ 0.49 $ & $ 0.46 $ & $ 0.51 $ & $ 0.51 $ & $ 0.55 $ & $ 0.49 $
& &  
$ 15.15 $ & $ 8.08 $ & $ 6.21 $ & $ 5.84 $ & $ 7.69 $ & $ 8.59 $
\\   
%& \\ 
\multicolumn{15}{c}{$ n_{r} = 1000 $, Weibull} \\ 
1 FPC      &  
$ 1.29 $ & $ 1.90 $ & $ 0.41 $ & $ 0.40 $ & $ 0.87 $ & $ 1.24 $ & $ 1.02 $
& &  
$ 56.84 $ & $ 18.51 $ & $ 6.53 $ & $ 8.35 $ & $ 36.29 $ & $ 25.30 $     
\\         
2 FPCs      &  
$ 0.37 $ & $ 0.60 $ & $ 0.32 $ & $ 0.32 $ & $ 0.40 $ & $ 0.44 $ & $ 0.41 $            
& &  
$ 12.91 $ & $ 7.48 $ & $ 5.58 $ & $ 4.86 $ & $ 6.55 $ & $ 7.47 $          
\\    
Adaptive      &  
$ 0.37 $ & $ 0.59 $ & $ 0.32 $ & $ 0.32 $ & $ 0.40 $ & $ 0.44 $ & $ 0.41 $
& &  
$ 12.91 $ & $ 7.29 $ & $ 5.48 $ & $ 4.86 $ & $ 6.57 $ & $ 7.42 $        
\\                             
Rich      &  
$ 0.40 $ & $ 0.40 $ & $ 0.30 $ & $ 0.31 $ & $ 0.35 $ & $ 0.42 $ & $ 0.36 $      
& &  
$ 12.08 $ & $ 6.39 $ & $ 5.11 $ & $ 4.89 $ & $ 5.63 $ & $ 6.82 $         
\\
K\&U 1      &  
$ 0.26 $ & $ 1.19 $ & $ 4.31 $ & $ 2.29 $ & $ 4.41 $ & $ 1.24 $ & $ 2.28 $       
& &  
$ 88.26 $ & $ 20.10 $ & $ 19.14 $ & $ 43.17 $ & $ 96.79 $ & $ 53.49 $               
\\ 
K\&U 2      &  
$ 0.36 $ & $ 0.51 $ & $ 0.62 $ & $ 0.34 $ & $ 0.54 $ & $ 1.04 $ & $ 0.57 $      
& &  
$ 22.56 $ & $ 7.70 $ & $ 5.34 $ & $ 5.35 $ & $ 11.99 $ & $ 10.59 $   
\\       
NP      &  
$ 0.48 $ & $ 0.57 $ & $ 0.57 $ & $ 0.61 $ & $ 0.59 $ & $ 0.67 $ & $ 0.58 $
& &  
$ 15.05 $ & $ 7.94 $ & $ 6.13 $ & $ 5.97 $ & $ 7.62 $ & $ 8.54 $       
\\   
%& \\ 
\multicolumn{15}{c}{$ n_{r} = 500 $, normal mixture} \\ 
1 FPC      &  
$ 0.28 $ & $ 0.36 $ & $ 0.74 $ & $ 0.78 $ & $ 1.09 $ & $ 0.76 $ & $ 0.67 $
& &  
$ 62.96 $ & $ 31.63 $ & $ 16.68 $ & $ 10.82 $ & $ 14.20 $ & $ 27.26 $
\\         
2 FPCs      &  
$ 0.27 $ & $ 0.23 $ & $ 0.31 $ & $ 0.39 $ & $ 0.26 $ & $ 0.36 $ & $ 0.30 $
& &  
$ 9.76 $ & $ 7.47 $ & $ 7.22 $ & $ 7.74 $ & $ 11.84 $ & $ 8.81 $
\\    
3 FPCs      &  
$ 0.33 $ & $ 0.30 $ & $ 0.40 $ & $ 0.36 $ & $ 0.31 $ & $ 0.41 $ & $ 0.35 $
& &  
$ 9.94 $ & $ 7.62 $ & $ 8.00 $ & $ 8.49 $ & $ 11.92 $ & $ 9.20 $
\\         
4 FPCs      &  
$ 0.39 $ & $ 0.37 $ & $ 0.42 $ & $ 0.36 $ & $ 0.35 $ & $ 0.44 $ & $ 0.39 $
& &  
$ 10.38 $ & $ 7.78 $ & $ 8.27 $ & $ 8.78 $ & $ 12.29 $ & $ 9.50 $
\\    
Adaptive      &  
$ 0.27 $ & $ 0.23 $ & $ 0.32 $ & $ 0.39 $ & $ 0.27 $ & $ 0.36 $ & $ 0.31 $
& &  
$ 9.95 $ & $ 7.59 $ & $ 7.33 $ & $ 7.77 $ & $ 11.84 $ & $ 8.90 $
\\                             
Rich      &  
$ 0.27 $ & $ 0.25 $ & $ 0.32 $ & $ 0.29 $ & $ 0.26 $ & $ 0.31 $ & $ 0.28 $
& &  
$ 9.07 $ & $ 7.01 $ & $ 6.96 $ & $ 7.72 $ & $ 11.11 $ & $ 8.37 $
\\
K\&U 1      &  
$ 0.30 $ & $ 0.24 $ & $ 1.10 $ & $ 0.62 $ & $ 0.64 $ & $ 1.22 $ & $ 0.69 $
& &  
$ 48.23 $ & $ 20.27 $ & $ 10.25 $ & $ 8.60 $ & $ 27.80 $ & $ 23.03 $
\\ 
K\&U 2      &  
$ 0.34 $ & $ 0.27 $ & $ 0.61 $ & $ 0.43 $ & $ 0.38 $ & $ 0.63 $ & $ 0.44 $
& &  
$ 17.03 $ & $ 8.79 $ & $ 7.85 $ & $ 8.79 $ & $ 21.22 $ & $ 12.74 $
\\       
K\&U 3      &  
$ 0.33 $ & $ 0.32 $ & $ 0.46 $ & $ 0.40 $ & $ 0.37 $ & $ 0.52 $ & $ 0.40 $
& &  
$ 14.79 $ & $ 7.76 $ & $ 7.26 $ & $ 8.47 $ & $ 16.17 $ & $ 10.89 $
\\       
K\&U 4      &  
$ 0.37 $ & $ 0.37 $ & $ 0.45 $ & $ 0.39 $ & $ 0.39 $ & $ 0.52 $ & $ 0.41 $
& &  
$ 14.33 $ & $ 7.58 $ & $ 7.21 $ & $ 8.44 $ & $ 15.40 $ & $ 10.59 $
\\       
NP      &  
$ 0.39 $ & $ 0.40 $ & $ 0.46 $ & $ 0.39 $ & $ 0.39 $ & $ 0.53 $ & $ 0.43 $
& &  
$ 12.15 $ & $ 8.48 $ & $ 8.67 $ & $ 9.53 $ & $ 14.08 $ & $ 10.58 $
\\   
%& \\ 
\multicolumn{15}{c}{$ n_{r} = 1000 $, normal mixture} \\ 
1 FPC      &  
$ 0.45 $ & $ 0.59 $ & $ 1.31 $ & $ 1.28 $ & $ 2.07 $ & $ 1.20 $ & $ 1.15 $
& &  
$ 119.08 $ & $ 56.71 $ & $ 25.46 $ & $ 12.27 $ & $ 15.54 $ & $ 45.81 $
\\         
2 FPCs      &  
$ 0.38 $ & $ 0.27 $ & $ 0.38 $ & $ 0.54 $ & $ 0.30 $ & $ 0.42 $ & $ 0.38 $
& &  
$ 9.99 $ & $ 8.14 $ & $ 8.09 $ & $ 8.36 $ & $ 12.12 $ & $ 9.34 $
\\    
3 FPCs      &  
$ 0.40 $ & $ 0.32 $ & $ 0.42 $ & $ 0.42 $ & $ 0.34 $ & $ 0.46 $ & $ 0.39 $
& &  
$ 10.06 $ & $ 7.76 $ & $ 8.19 $ & $ 8.68 $ & $ 11.69 $ & $ 9.28 $
\\         
4 FPCs      &  
$ 0.41 $ & $ 0.40 $ & $ 0.43 $ & $ 0.39 $ & $ 0.37 $ & $ 0.50 $ & $ 0.42 $
& &  
$ 10.20 $ & $ 7.79 $ & $ 8.22 $ & $ 8.76 $ & $ 11.80 $ & $ 9.36 $
\\    
Adaptive      &  
$ 0.38 $ & $ 0.27 $ & $ 0.38 $ & $ 0.54 $ & $ 0.30 $ & $ 0.42 $ & $ 0.38 $
& &  
$ 9.97 $ & $ 8.12 $ & $ 8.08 $ & $ 8.35 $ & $ 12.10 $ & $ 9.33 $
\\
Rich      &  
$ 0.36 $ & $ 0.29 $ & $ 0.39 $ & $ 0.34 $ & $ 0.29 $ & $ 0.35 $ & $ 0.34 $
& &  
$ 9.51 $ & $ 7.40 $ & $ 7.06 $ & $ 8.09 $ & $ 10.96 $ & $ 8.60 $
\\
K\&U 1      &  
$ 0.51 $ & $ 0.40 $ & $ 2.00 $ & $ 0.97 $ & $ 1.28 $ & $ 2.06 $ & $ 1.20 $
& &  
$ 86.96 $ & $ 35.82 $ & $ 14.78 $ & $ 9.26 $ & $ 40.05 $ & $ 37.37 $
\\ 
K\&U 2      &  
$ 0.48 $ & $ 0.32 $ & $ 1.01 $ & $ 0.57 $ & $ 0.48 $ & $ 0.84 $ & $ 0.62 $
& &  
$ 18.94 $ & $ 10.26 $ & $ 9.38 $ & $ 9.57 $ & $ 29.15 $ & $ 15.46 $
\\       
K\&U 3      &  
$ 0.38 $ & $ 0.37 $ & $ 0.55 $ & $ 0.46 $ & $ 0.44 $ & $ 0.65 $ & $ 0.48 $
& &  
$ 16.31 $ & $ 8.16 $ & $ 7.57 $ & $ 8.80 $ & $ 16.85 $ & $ 11.54 $
\\       
K\&U 4      &  
$ 0.43 $ & $ 0.43 $ & $ 0.53 $ & $ 0.46 $ & $ 0.46 $ & $ 0.65 $ & $ 0.49 $
& &  
$ 15.61 $ & $ 7.96 $ & $ 7.47 $ & $ 8.76 $ & $ 16.17 $ & $ 11.19 $
\\       
NP      &  
$ 0.47 $ & $ 0.48 $ & $ 0.53 $ & $ 0.46 $ & $ 0.46 $ & $ 0.66 $ & $ 0.51 $
& &  
$ 12.37 $ & $ 8.65 $ & $ 8.77 $ & $ 9.57 $ & $ 13.94 $ & $ 10.66 $
\\   
\end{tabular}
\end{adjustbox}
\label{nonDRM_combined_density_quan} 
\end{table}

%%%%%%%%%%%%%%%%%%%%%%%%%%%%%%%%%%%%%%%%%%%%%%%%%%%%%%%%%%%%%%%%%%%%%%

\newpage 

\bibliographystyle{abbrvnat}
\bibliography{Archer_proposal}

\end{document}